\documentclass[]{pasj02} 
\usepackage[switch,mathlines]{lineno} 

\usepackage{xcolor}
\jyear{2025}
\Received{2025/02/27}
\Accepted{2025/07/22}

\usepackage{comment}

\newcommand{\addr}[1]{\color{red}{#1 }\color{black}}

\newcommand{\iac}{Instituto de Astrof\'\i sica de Canarias (IAC), 38205 La Laguna, Tenerife, Spain}
\newcommand{\komaba}{Department of Multi-Disciplinary Sciences, The University of Tokyo, 3-8-1 Komaba, Meguro, Tokyo 153-8902, Japan}
\newcommand{\komabains}{Komaba Institute for Science, The University of Tokyo, 3-8-1 Komaba, Meguro, Tokyo 153-8902, Japan}
\newcommand{\abc}{Astrobiology Center, 2-21-1 Osawa, Mitaka, Tokyo 181-8588, Japan}
\newcommand{\naoj}{National Astronomical Observatory of Japan, 2-21-1 Osawa, Mitaka, Tokyo 181-8588, Japan}
\newcommand{\astron}{Department of Astronomy, The University of Tokyo, 7-3-1 Hongo, Bunkyo, Tokyo 113-0033, Japan}
\newcommand{\sokendai}{Department of Astronomical Science, The Graduate University for Advanced Studies (SOKENDAI), 2-21-1 Osawa, Mitaka, Tokyo 181-8588, Japan}
\newcommand{\ep}{Department of Earth and Planetary Science, The University of Tokyo, 7-3-1 Hongo, Bunkyo, Tokyo 113-0033, Japan}
\newcommand{\utops}{UTokyo Organization for Planetary and Space Science,The University of Tokyo, 7-3-1 Hongo, Bunkyo, Tokyo 113-0033, Japan}
\newcommand{\ritsumei}{Department of Physical Sciences, Ritsumeikan University, 
1-1-1 Noji-higashi, Kusatsu, Shiga 525-8577, Japan}
\newcommand{\subaru}{Subaru Telescope, National Astronomical Observatory of Japan, 650 N. Aohoku Place, Hilo, HI 96720, USA}

\newcommand{\elsi}{Earth-Life Science Institute (ELSI), Institute of Science Tokyo, 2-12-1 Ookayama, Meguro, Tokyo 152-8551, Japan}

\newcommand{\isas}{Institute of Space and Astronautical Science, Japan Aerospace Exploration Agency, 3-1-1 Yoshinodai, Chuo, Sagamihara, Kanagawa, 252-5210, Japan}

\newcommand{\lagna}{Departamento de Astrof\'{i}sica, Universidad de La Laguna (ULL), E-38206 La Laguna, Tenerife, Spain}

\newcommand{\gron}{Kapteyn Astronomical Institute, University of Groningen, P.O. Box 800, 9700 AV Groningen, Netherlands}

\newcommand{\sun}{\solar}

\begin{document} 
\title{The mass of TOI-654~b: A short-period sub-Neptune transiting a mid-M dwarf}

\author{
 Kai \textsc{Ikuta},\altaffilmark{1}\altemailmark\orcid{0000-0002-5978-057X} 
 Norio \textsc{Narita},\altaffilmark{2,3,4}\orcid{0000-0001-8511-2981}
 Takuya \textsc{Takarada},\altaffilmark{4,5}\orcid{0009-0006-9082-9171}
 Teruyuki \textsc{Hirano},\altaffilmark{4,5,6}\orcid{0000-0003-3618-7535}
  Akihiko \textsc{Fukui},\altaffilmark{2,3}\orcid{0000-0002-4909-5763}
  Hiroyuki Tako \textsc{Ishikawa},\altaffilmark{7}\orcid{0000-0001-6309-4380}
  Yasunori \textsc{Hori},\altaffilmark{4,5,6}\orcid{0000-0003-4676-0251}
  Tadahiro \textsc{Kimura},\altaffilmark{8,9}\orcid{0000-0001-8477-2523}
  Takanori \textsc{Kodama},\altaffilmark{10}\orcid{0000-0001-9032-5826}
 Masahiro \textsc{Ikoma},\altaffilmark{5,11}\orcid{0000-0002-5658-5971} 
 Jerome P. \textsc{de Leon},\altaffilmark{1}\orcid{0000-0002-6424-3410}
 Kiyoe \textsc{Kawauchi},\altaffilmark{12}\orcid{0000-0003-1205-5108}
 Masayuki \textsc{Kuzuhara},\altaffilmark{4,5,6}\orcid{0000-0002-4677-9182}
 Gaia \textsc{Lacedelli},\altaffilmark{3}\orcid{0000-0002-4197-7374}
 John H. \textsc{Livingston},\altaffilmark{4,5,6}\orcid{0000-0002-4881-3620}
 Mayuko \textsc{Mori},\altaffilmark{4,5}\orcid{0000-0003-1368-6593}
 Felipe \textsc{Murgas},\altaffilmark{3,13}\orcid{0000-0001-9087-1245}
 Enric \textsc{Palle},\altaffilmark{3,13}\orcid{0000-0003-0987-1593}
 Hannu \textsc{Parviainen},\altaffilmark{3,13}\orcid{0000-0001-5519-1391}
   Noriharu \textsc{Watanabe},\altaffilmark{1}\orcid{0000-0002-7522-8195}
    Izuru \textsc{Fukuda},\altaffilmark{1}\orcid{0000-0002-9436-2891}
    Hiroki \textsc{Harakawa},\altaffilmark{14}\orcid{0000-0002-7972-0216}
    Yuya \textsc{Hayashi},\altaffilmark{1}\orcid{0000-0001-8877-0242}
    Klaus \textsc{Hodapp},\altaffilmark{15}\orcid{0000-0003-0786-2140}
    Keisuke \textsc{Isogai},\altaffilmark{1,16}\orcid{0000-0002-6480-3799}  
    Taiki \textsc{Kagetani},\altaffilmark{1,5}\orcid{0000-0002-5331-6637}
    Yugo \textsc{Kawai},\altaffilmark{1}\orcid{0000-0002-0488-6297}
    Vigneshwaran \textsc{Krishnamurthy},\altaffilmark{17}\orcid{0000-0003-2310-9415}
    Tomoyuki \textsc{Kudo},\altaffilmark{14}\orcid{0000-0002-9294-1793}
    Takashi \textsc{Kurokawa},\altaffilmark{5,18}
    Nobuhiko \textsc{Kusakabe},\altaffilmark{4,5}\orcid{0000-0001-9194-1268}
    Jun \textsc{Nishikawa},\altaffilmark{5,6,4}\orcid{0000-0001-9326-8134}
    Stevanus K. \textsc{Nugroho},\altaffilmark{4,5}\orcid{0000-0003-4698-6285}
    Masashi \textsc{Omiya},\altaffilmark{4,5}
    Takuma \textsc{Serizawa},\altaffilmark{5,18} \orcid{0009-0009-5823-0793}
    Aoi \textsc{Takahashi},\altaffilmark{19} \orcid{0000-0003-3881-3202}
    Huan-Yu \textsc{Teng},\altaffilmark{20}\orcid{0000-0003-3860-6297}
    Yuka \textsc{Terada},\altaffilmark{21,22}\orcid{0000-0003-2887-6381}
    Akitoshi \textsc{Ueda},\altaffilmark{4,5,6}
    S\'{e}bastien \textsc{Vievard},\altaffilmark{14}\orcid{0000-0003-4018-2569}
    Yujie \textsc{Zou},\altaffilmark{1}\orcid{0000-0002-5609-4427}
    Takayuki \textsc{Kotani},\altaffilmark{4,5,6}\orcid{0000-0001-6181-3142}
    and 
    Motohide \textsc{Tamura} \altaffilmark{23,4,5}\orcid{0000-0002-6510-0681}
}

\altaffiltext{1}{\komaba}
\altaffiltext{2}{\komabains}
\altaffiltext{3}{\iac}
\altaffiltext{4}{\abc}
\altaffiltext{5}{\naoj}
\altaffiltext{6}{\sokendai}
\altaffiltext{7}{Department of Physics and Astronomy, The University of Western Ontario, 1151 Richmond St, London, Ontario, N6A 3K7, Canada}
\altaffiltext{8}{\utops}
\altaffiltext{9}{\gron}
\altaffiltext{10}{\elsi}
\altaffiltext{11}{\ep}
\altaffiltext{12}{\ritsumei}
\altaffiltext{13}{\lagna}
\altaffiltext{14}{\subaru}
\altaffiltext{15}{University of Hawaii, Institute for Astronomy, 640 N. Aohoku Place, Hilo, HI 96720, USA}
\altaffiltext{16}{Okayama Observatory, Kyoto University, 3037-5 Honjo, Kamogata, Asakuchi, Okayama 719-0232, Japan}
\altaffiltext{17}{Trottier Space Institute at McGill, McGill University, 3550 University Street, Montreal, QC H3A 2A7, Canada}
\altaffiltext{18}{Institute of Engineering, Tokyo University of Agriculture and Technology, 2-24-26 Nakacho, Koganei, Tokyo, 184-8588, Japan}
\altaffiltext{19}{\isas}
\altaffiltext{20}{CAS Key Laboratory of Optical Astronomy, National Astronomical Observatories, Chinese Academy of Sciences, Beijing 100101, China}
\altaffiltext{21}{Institute of Astronomy and Astrophysics, Academia Sinica, P.O. Box 23-141, Taipei 10617, Taiwan, R.O.C.}
\altaffiltext{22}{Department of Astrophysics, National Taiwan University, Taipei 10617, Taiwan, R.O.C.}
\altaffiltext{23}{\astron}

\email{kaiikuta.astron@gmail.com}



\maketitle

\begin{abstract}
Sub-Neptunes are small planets between the size of the Earth and Neptune. The orbital and bulk properties of transiting sub-Neptunes can provide clues for their formation and evolution of small planets. 
In this paper, we report on follow-up observations of a planetary system around the mid-M dwarf TOI-654, whose transiting sub-Neptune TOI-654~b ($P=1.53$ day) is validated as a suitable target for the atmospheric observation. 
We measure the planetary mass and stellar properties with the InfraRed Doppler instrument (IRD) mounted on the Subaru telescope and obtain the stellar and planetary properties from additional transit observations by the Transit Exoplanetary Survey Satellite (TESS) and a series of the Multicolor Simultaneous Camera for studying Atmospheres of Transiting exoplanets (MuSCAT). 
As a result, the planetary mass of TOI-654~b is determined to be $M_{{\rm p}} = 8.71  \pm 1.25 M_{\earth}$, and the radius is updated to be $R_{\rm p} = 2.378 \pm  0.089 R_{\earth}$. The bulk density suggests that the planet is composed of a rocky and volatile-rich core or a rocky core surrounded by a small amount of H/He envelope. 
TOI-654~b is one of unique planets located around the radius valley and and also on the outer edge of the Neptune desert.
The precise mass determination enables us to constrain the atmospheric properties with future spectroscopic observations especially for the emission by the James Webb Space Telescope and Ariel.
\end{abstract}


\section{Introduction} \label{sec:intro}
Recent dedicated photometric surveys by space telescopes such as Kepler \citep{Borucki11} and Transiting Exoplanet Survey Satellite (TESS; \cite{Ricker15}) have facilitated the discovery of many transiting planets and revealed that there are ubiquitously small planets with sizes between the Earth and Neptune (e.g., \cite{Batalha13}; \cite{Bryson21}).
The radius distribution of small planets exhibits a bimodality with a valley around a radius of 1.8 $R_{\earth}$, and the planets with the radii below and above the valley are respectively classified as Super-Earths and Sub-Neptunes \citep{Fulton17,Fulton18, Hirano18, Eylen18,Cloutier_Menou20,Cloutier20,Petigura22}.
The bimodality of the planetary radius has been interpreted as the consequences of either of the following mechanisms. The valley can be majorly explained by the mass-loss of the primordial atmosphere by X-ray and ultraviolet (XUV) radiations from the host star \citep{Owen13,Lopez13,Owen17,Lopez18} or heating from the cooling core of the planet \citep{Ginzburg18,Gupta19,Gupta21,Gupta22}.
Both of the processes have been demonstrated (e.g., \cite{Estrela20,Rogers21,Ho23}) as the same mechanism determined by the location of the Bondi radius \citep{Owen24}. Atmospheric loss can be also provoked with a boil-off by the stellar XUV after the stellar accretion disk disperses (e.g., \cite{Ikoma12, Rogers24}) and giant impacts (e.g., \cite{Inamdar15,Liu15}).
Alternatively, the valley can be explained by other various processes such as gas-poor formation (e.g., \cite{Lee14,Lopez18}) or migration of planets with ice rich cores after the formation beyond snow lines (e.g., \cite{Venturini20,Burn24}).

The location of the radius valley has been investigated in terms of the dependency especially on the stellar mass \citep{Cloutier_Menou20,Eylen21,Petigura22,Luque22,Venturini24,Ho24,Parc24,Gaidos24}.
However, the existence for M dwarfs is under debate because a limited number of small planets on the radius valley has been confirmed (e.g., \cite{Cloutier20, Alvarez23, Hannu24, Lacedelli24}).
Thus, the detailed characterization of additional sub-Neptunes is required to explore the formation and evolution of small planets around the radius valley of M dwarfs.

In this paper, we report on the characterization of a short-period sub-Neptune around the mid-M dwarf TOI-654 (Table \ref{tb:stellar}), whose transiting planet TOI-654~b is validated in \citet{Hord24} as a suitable target for the transmission and emission spectroscopy with the James Webb Space Telescope (JWST; \cite{jwst}) and Ariel \citep{ariel}. 
We characterize the system by follow-up observations with the InfraRed Doppler (IRD) instrument on Subaru telescope \citep{Tamura12,Kotani18,Kuzuhara18} 
and a series of the Multicolor Simultaneous Camera for studying Atmospheres of Transiting exoplanets (MuSCAT; \cite{muscat}) to determine the stellar and planetary properties especially for the planetary mass.
This paper is a part of the intensive programs with the Subaru/IRD, dedicated for follow-up observations of exoplanets (e.g., \cite{Hirano20b,Hirano21,Fukui22,Mori22,Kawauchi22,Kagetani23,Hirano23,Barkaoui24,Hori23}).
The rest of this paper is organized as follows. In Section \ref{sec:data}, we describe the data of the photometry for the planetary transits and high resolution spectroscopy for the stellar characterization and radial velocity.
In Section \ref{sec:result}, we derive the stellar and planetary properties both from photometry and spectroscopy.
In Section \ref{sec:discussion}, we discuss the planetary composition and future atmospheric observations. In Section \ref{sec:conclusion}, we conclude this paper. In Appendix \ref{sec:activity_analysis}, we exhibit photometric light curves and their period analysis. In Appendix \ref{sec:appendix}, we describe other derived parameters.

\begin{table}[tb]
\caption{Stellar parameters of TOI-654}
\begin{center}
\begin{tabular}{lc}
Parameter & TOI-654 \\ \hline
(Literature Values) &  \\
TIC  & 35009898  \\
2MASS ID   & J10585379+0532468  \\
Gaia ID &  3788670679927991296 \\ 
$\alpha$ (J2000)$^{\rm a}$  &10:58:53.90  \\
$\delta$ (J2000)$^{\rm a}$  &  -05:32:51.30 \\
$\mu_\alpha \cos \delta$ (mas yr$^{-1}$)$^{\rm a}$     &99.296 $\pm$ 0.022  \\
$\mu_\delta$ (mas yr$^{-1}$)$^{\rm a}$   & -266.053 $\pm$ 0.017 \\
parallax (mas)$^{\rm a}$ &   17.2861 $\pm$ 0.0165  \\
Gaia $G$ (mag)$^{\rm a}$  & 13.4131 $\pm$ 0.0006  \\
TESS $T$ (mag)$^{\rm b}$   &12.2255 $\pm$ 0.0074  \\
$V$ (mag)$^{\rm b}$ & 14.54 $\pm$ 0.010 \\
$J$ (mag)$^{\rm b}$   & 10.744 $\pm$ 0.024 \\
$H$ (mag)$^{\rm b}$   & 10.139 $\pm$ 0.022  \\
$K$ (mag)$^{\rm b}$  & 9.918 $\pm$ 0.021  \\ \hline
(Derived Values) &   \\
Spectral type & M2V \\
$d$ (pc)$^{\rm c}$   & $57.851^{+0.055}_{-0.056}$ \\
$T_{\rm eff}$ (K)$^{\rm c}$   & $3521^{+33}_{-48}$  \\
$\lbrack{\rm Na/H}\rbrack$ (dex)  & 0.11 $\pm$ 0.20  \\
$\lbrack{\rm Mg/H}\rbrack$ (dex)  & 0.24 $\pm$ 0.35  \\
$\lbrack{\rm Ca/H}\rbrack$ (dex) & 0.15 $\pm$ 0.24   \\
$\lbrack{\rm Ti/H}\rbrack$ (dex) & 0.45 $\pm$ 0.22   \\
$\lbrack{\rm Cr/H}\rbrack$ (dex) & 0.12 $\pm$ 0.13  \\
$\lbrack{\rm Mn/H}\rbrack$ (dex) & 0.16 $\pm$ 0.27  \\
$\lbrack{\rm Fe/H}\rbrack$ (dex) &0.12 $\pm$ 0.18   \\
$\lbrack{\rm Sr/H}\rbrack$ (dex) &  0.22 $\pm$ 0.36  \\
$\lbrack{\rm M/H}\rbrack$ (dex) & 0.17 $\pm$ 0.07  \\
$\log g$ (cgs)  & $4.793^{+0.037}_{ -0.036}$  \\
$M_{\rm s}$ ($M_{\sun}$)   & $0.419 \pm 0.009$  \\ 
$R_{\rm s}$ ($R_{\sun}$)  & $0.430 \pm 0.013$   \\ 
$\rho_{\rm s}$ (g cm$^{-3}$) & $7.43^{ +0.91}_{-0.80}$  \\ 
$L_{\rm s}$ ($L_{\sun}$)  & $0.0246\pm 0.0005$  \\  
$v \sin i_{\rm s}$ (km s$^{-1}$)  & < 3  \\ \hline
\end{tabular} \label{tb:stellar}
\end{center}
\begin{tabnote}
\par\noindent
\footnotemark[a] Gaia DR3 \citep{Gaia}
\par\noindent
\footnotemark[b] TESS Input Catalog v8.2 \citep{Stassun19}
\par\noindent
\footnotemark[c] The distance and stellar effective temperature are derived from the SED (Section \ref{sec:sed}). We note that the uncertainty of the stellar effective temperature does not take into account an inherent uncertainty in the stellar atmospheric model.
\end{tabnote}
\end{table}

\begin{table}[tb]
\caption{Transit Observations of TOI-654~b}
\begin{center}
\begin{tabular}{lc}
Date & Usage \\ \hline
(TESS)  & \\
2019/02/28 - 03/26 (Sector 9) & (a,b)  \\
2021/03/07 - 04/02 (Sector 36) & (a,b) \\
2021/11/06 - 12/02 (Sector 45) & (a,b)  \\
2021/12/02 - 12/30 (Sector 46) & (a,b)  \\
2023/02/12 - 03/10 (Sector 62) & (b)  \\
2023/11/11 - 12/07 (Sector 72) & (b)  \\ \hline
(MuSCAT2) &  \\
2019/05/20 & Partial transit \\
2019/05/23 & (a) \\
2020/01/13  & Nearby Moon \\
2020/02/03 & Partial transit  \\
2020/02/05 & (a) \\
2021/03/02 & Clouds \\
2021/03/19 & (b) \\ \hline
(MuSCAT3) & \\ 
2021/01/28 & (b) \\ \hline
\end{tabular} \label{tb:transit_obs}
\end{center}
\begin{tabnote}
\par\noindent
\footnotemark[(a)] \citet{Hord24}
\par\noindent
\footnotemark[(b)] This study
\end{tabnote}
\end{table}

\section{Observations} \label{sec:data}
\subsection{TESS photometry} \label{sec:tess}

TOI-654 (TIC 35009898) was observed by the TESS with two-minute cadence in its sectors 9, 36, 45, 46, 62, and 72.
The signal of a transiting planetary candidate with a period of 1.528 days (TOI-654.01) was identified by the Science Processing Operations Center (SPOC) pipeline \citep{Jenkins16} from the data in Sector 9 \citep{Guerrero21}. \citet{Hord24} statistically validate TOI-654.01 from the first four sectors (9, 36, 45, and 46) with the Gaia DR3 catalog \citep{Gaia} and the TOI catalog \citep{Guerrero21} by incorporating follow-up observations with ground-based reconnaissance photometry through the TESS Follow-up Observing Program (TFOP; \cite{Collins19}).

We extract the Presearch Data Conditioning Simple Aperture Photometry (PDC-SAP) for six sectors (9, 36, 45, 46, 62, and 72) from the Mikulski Archive for Space Telescopes (MAST). 
We remove data with bad-quality flags for each sector and use the data within transit windows of three times the predicted transit duration (= 67 min) (Figure \ref{fig:tess}). The data is normalized by the data out of the transit window for each sector because there is no flux variability attributed to stellar activity or systematics (Appendix \ref{sec:activity_analysis}).

\subsection{Ground-based photometry -- TCS 1.52m / MuSCAT2} \label{sec:m2}
We conducted follow-up transit observations of TOI-654~b seven times from 2019 to 2021 with the multi-band simultaneous camera MuSCAT2 \citep{muscat2}, mounted on the 1.52m Carlos S\'{a}nchez telescope (TCS) at the Teide Observatory in the Canary Islands, Spain (Table \ref{tb:transit_obs}). MuSCAT2 has four optical channels, each of which is equipped with a 1k $\times$ 1k CCD camera with a pixel scale of 0".44 pixel$^{-1}$, and is capable of simultaneous imaging in $g$-, $r$-, $i$-, and $z_s$-bands with a field of view of 7'4 $\times$ 7'4. 
We only use the data with the best quality for the transit analysis, which was obtained on 2021 March 19 UTC from BJD 2459293.383 to 2459293.560 (Figure \ref{fig:m2_654}). 
The exposure times were set to be 30, 60, 60, and 45 s in $g$-, $r$-, $i$-, and $z_s$-bands, respectively.
We calibrate the images and perform aperture photometry to extract relative photometry by following the procedure described in \citet{Fukui16} and application to MuSCAT2 data (e.g., \cite{Fukui21}) using a dedicated pipeline developed in \citet{Fukui11}.

\subsection{Ground-based photometry -- FTN 2m / MuSCAT3} \label{sec:m3}
\begin{figure*}[tb]
\begin{center}
\includegraphics[width=\textwidth]{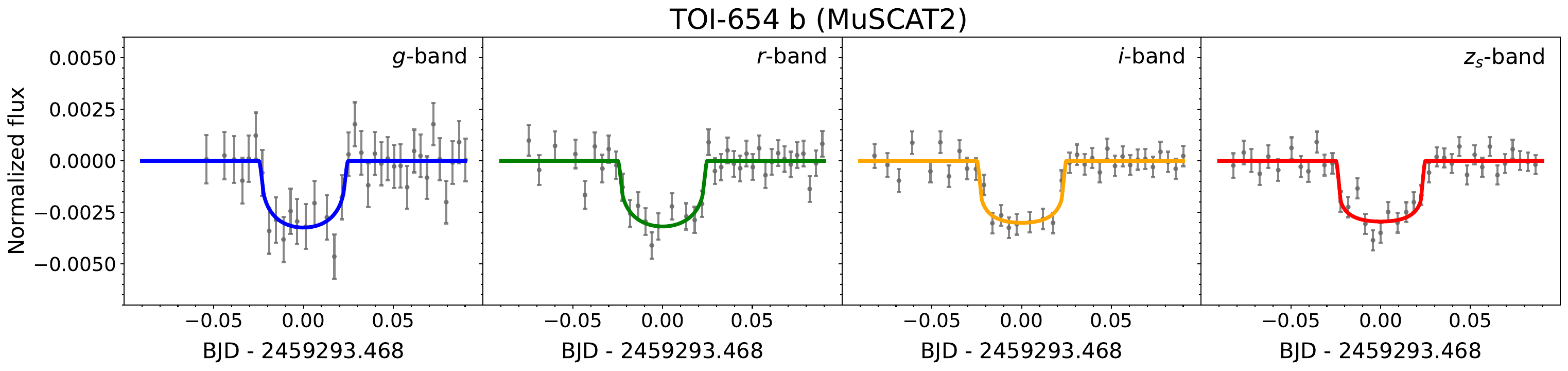}
\includegraphics[width=\textwidth]{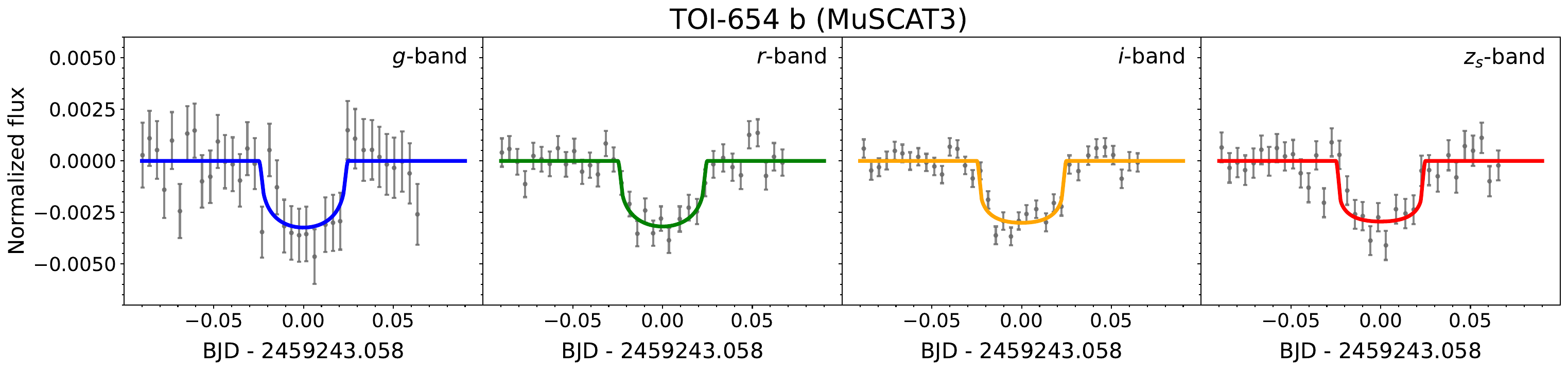}
\end{center}
\caption{Multicolor simultaneous light curves in $g$-, $r$-, $i$-, and $z_s$-bands, obtained by MuSCAT2 on 2021 March 19 (Top) and MuSCAT3 on 2021 January 28 (Bottom) (Section \ref{sec:m2} and \ref{sec:m3}). The data are jointly fitted with the transit model (blue, green, orange, and red) and baseline model, and the data are subtracted with the baseline models (gray). 
{Alt text: Time-series flux for the transits of the MuSCAT2 and MuSCAT3 data in $g$-, $r$-, $i$-, and $z_s$-bands, and the optimum models for each.} 
}
\label{fig:m2_654}
\end{figure*}

We conducted a follow-up transit observation of TOI-654~b on 2021 January 28 UTC from BJD 2459242.966 to 2459243.128 (Figure \ref{fig:m2_654}) with the multi-band simultaneous camera MuSCAT3 \citep{muscat3}, mounted on the 2m Faulkes Telescope North (FTN) at Haleakala Observatory in Hawaii, the United States. MuSCAT3 has four optical channels, each of which is equipped with a 2k$\times$2k CCD camera with a pixel scale of 0".266 pixel$^{-1}$, and is capable of simultaneous imaging in $g$-, $r$-, $i$-, and $z_s$-bands with a field of view of 9'1 $\times$ 9'1. The exposure times were set to be 30, 9, 15, and 13 s in $g$-, $r$-, $i$-, and $z_s$-bands, respectively.
The raw images are processed with \texttt{BANZAI} pipeline \citep{banzai}, and the aperture photometry is conducted in the same way for MuSCAT2 (Section \ref{sec:m2}).

\begin{figure}[tb]
\begin{center}
\includegraphics[width=8cm]{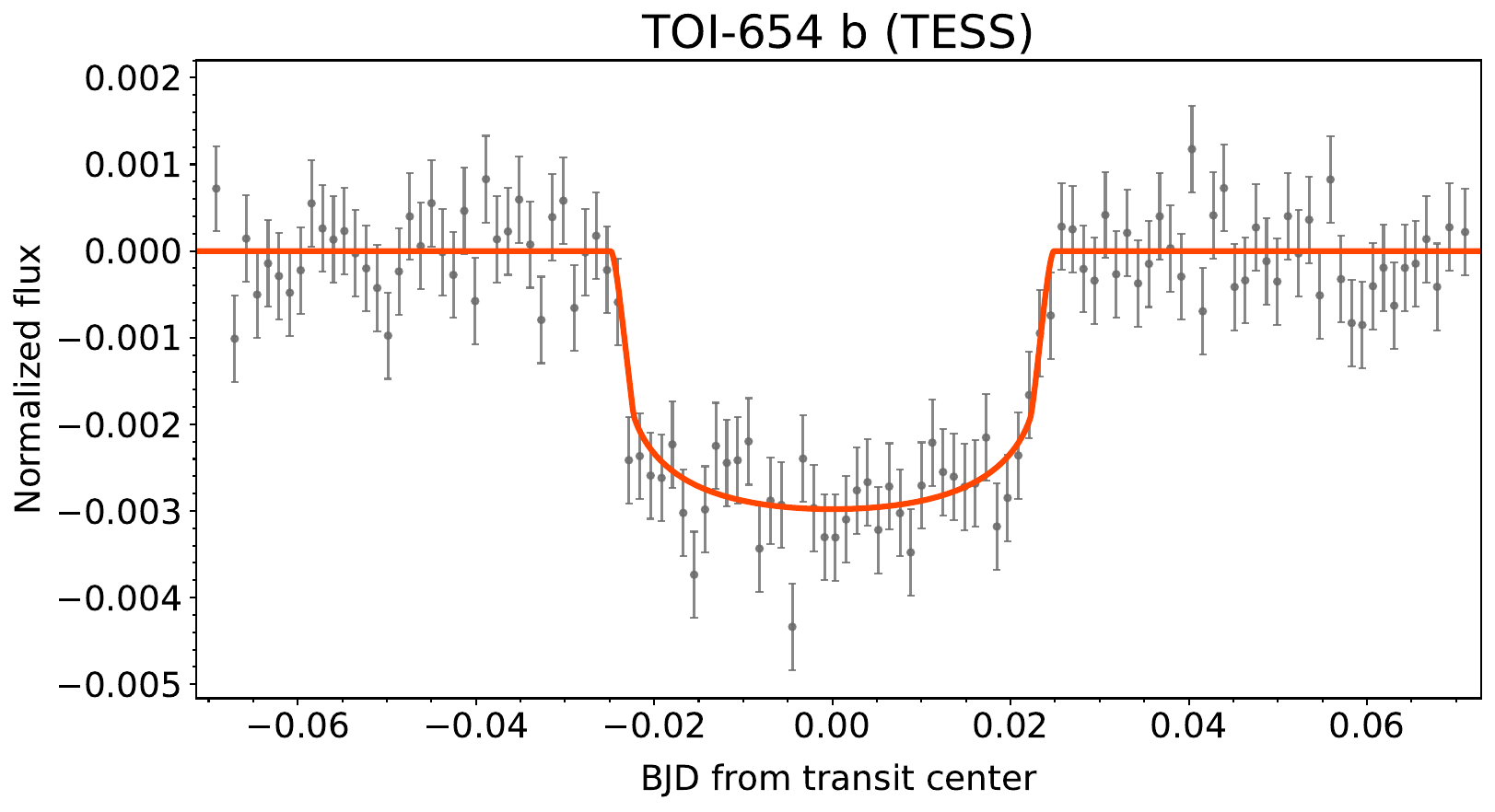}
\end{center}
\caption{Phase-folded TESS light curve (gray) with the derived orbital period (= 1.528 day) and optimum model (orange) near the transit center within transit windows of three times the transit duration of TOI-654~b (Section \ref{sec:tess}).
{Alt text: Time-series flux for the transit of the TESS data and the optimum model.} 
} \label{fig:tess}
\end{figure}

\begin{figure}[tb]
\begin{center}
\includegraphics[width=8cm]{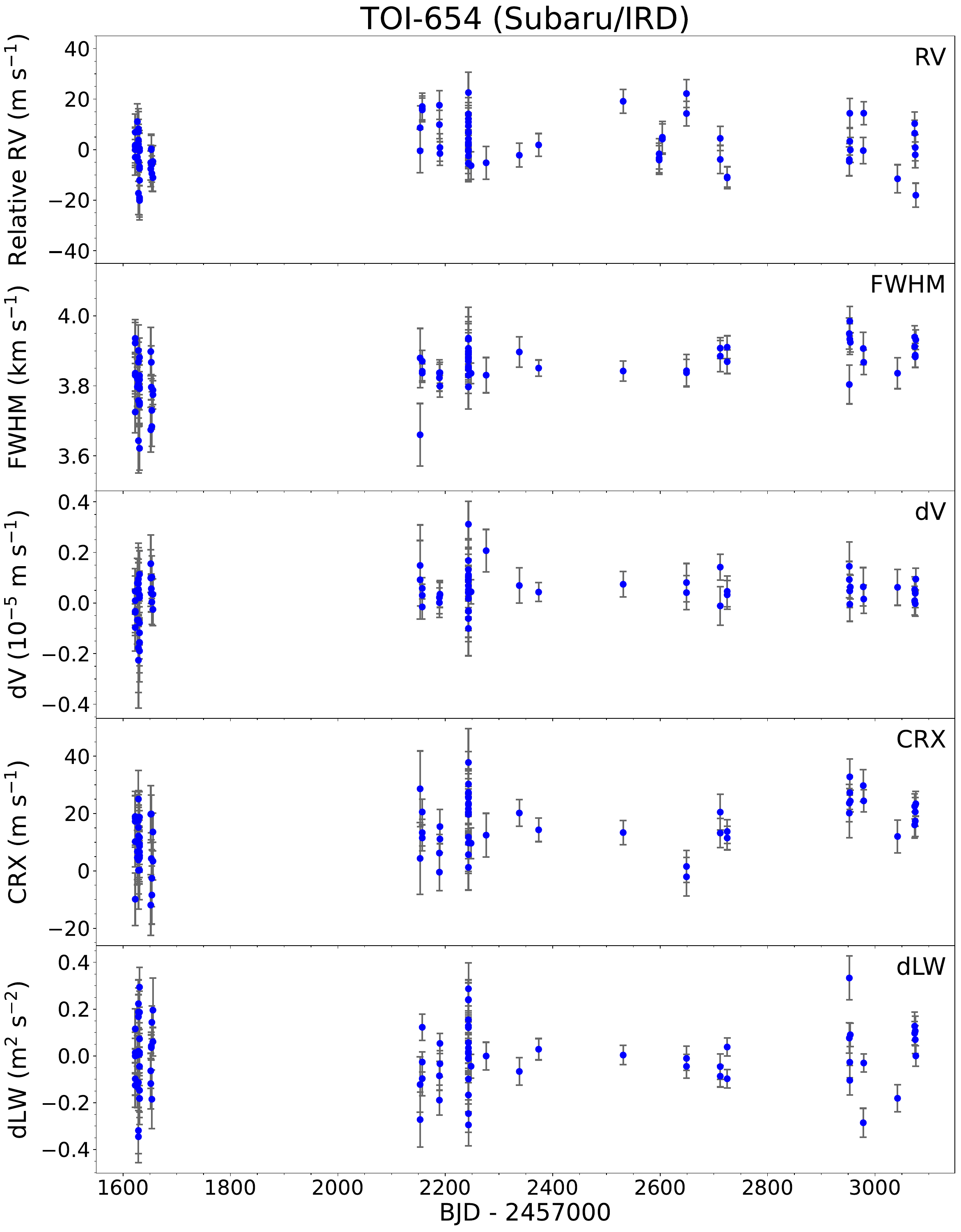}
\end{center}
\caption{Time-series RV, FWHM, dV, CRX, and dLW (blue) for TOI-654 from the IRD spectra (Section \ref{sec:ird}).
{Alt text: Time-series RV, FWHM, dV, CRX, and dLW, from the IRD spectra for each panel.} 
} \label{fig:rv}
\end{figure}

\subsection{High resolution spectroscopy -- Subaru 8.2m / IRD} \label{sec:ird}

We obtained time-series high-resolution spectra of TOI-654 with the
InfraRed Doppler (IRD) instrument \citep{Tamura12,Kotani18,Kuzuhara18}, mounted on the 8.2 m Subaru telescope from 2019 May 19 to 2023 May 11 UTC (BJD 2458622.800 to 2460075.891) under the Subaru/IRD intensive programs (ID: S19A-069I, S20B-088I, S21B-118I, and S23A-067I; PI: Norio Narita) (Figure \ref{fig:rv}). 
The IRD is a fiber-fed near-infrared spectrograph with the wavelength range of 930 to 1740 nm
and spectral resolution of $\sim70000$. 
The integration time per exposure was set to be from 300 to 1800 s, depending on the
observing condition for each night.  
We obtained 85 spectra with the signal-to-noise ratio from 18 to 72 per pixel at 1000 nm, and the raw data are reduced by the procedure in \citet{Kuzuhara18} and \citet{Hirano20}. 

We extract the radial velocity (RV) in the procedure of \citet{Hirano20} and \citet{Kuzuhara24}. 
Each observed spectrum is fitted with the stellar template spectrum processed from the individual spectra. The spectra are divided into segments, and the RV is calculated from each segment for each exposure. The typical internal precision and root mean square of the RV are 6.32 m s$^{-1}$ and 9.43 m s$^{-1}$, respectively.
The activity indicators, the full width at half maximum (FWHM; the line width), BiGauss (dV; the line asymmetry) by fitting Gaussian functions \citep{Santerne15}, chromatic index (CRX; the wavelength dependence of
RV) and the differential line width (dLW; the line width) \citep{Zechmeister18}, are also computed to investigate the stellar activity as in \citet{Harakawa22}.
We derive 80 data points for the activity indicators from the IRD spectra taken both in  $YJ$- and $H$-bands, and five data points only in the RV are taken only in $H$-band on the time from BJD 2459598.123 to 2459604.141 (Figure \ref{fig:rv}).

We note that we conducted the transmission spectroscopy of TOI-654~b from BJD 2459243.008 to 2459243.125 to investigate the atmospheric escape of Helium using the spectrum line of triplet He I (1083 nm) in the wavelength of the IRD spectra (e.g., \cite{Hirano20,Kawauchi22,Vignesh23,Hirano24}) at the same time of the transit observation with MuSCAT3 (Section \ref{sec:m3}). We report on the results of the atmospheric escape of TOI-654~b and other targets in a forthcoming paper (Kawauchi et al., in preparation).

\begin{figure}[tb]
\begin{center}
\includegraphics[width=8cm]{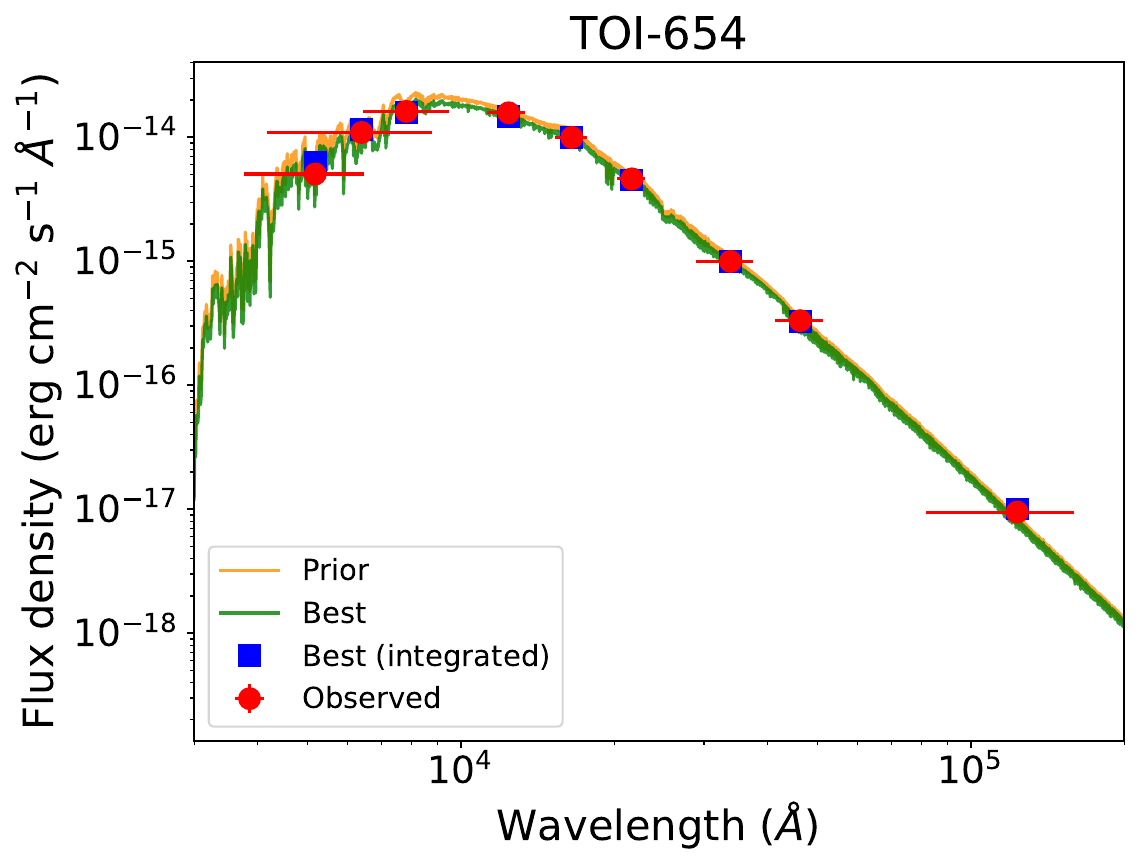}
\end{center}
\caption{Spectrum energy distributions (SED) of TOI-654. Yellow and green curves show the prior and posterior of the SED models. Red diamonds and blue squares are data and optimum, respectively (for details, Section \ref{sec:sed}).
{Alt text: Wavelength (\AA) versus flux density (erg cm$^{-2}$ s$^{-1}$ {\rm \AA}$^{-1})$} as the SED. Diamond and square show the observed data and optimum point. Two lines show the prior distribution for fitting the SED and the optimum model.
} \label{fig:SED}
\end{figure}

\section{Analyses and Results} \label{sec:result}

\subsection{Stellar properties}
\subsubsection{Spectroscopic properties} \label{sec:metal}
We analyze the IRD spectra to derive atmospheric parameters of TOI-654.
We use the template spectra extracted for the radial velocity, in which telluric features are removed and multiple frames are combined, because many parts of the original IRD spectra suffer from significant telluric features (both absorptions and airglow emissions).
The template spectra are subjected to the
analysis developed in \citet{Ishikawa20}.
The analysis is based on a line-by-line comparison between the equivalent widths (EWs) from observed spectra and those from synthetic spectra calculated with a one-dimensional LTE spectral code \citep{Tsuji78}.
We also interpolate the grid of MARCS models \citep{Gustafsson08} for the structure of the atmospheric layer.
The surface gravity $\log g$ is assumed to be the value calculated from mass and radius \citep{Stassun19} through the mass-$M_K$ and radius-$M_K$ relations respectively in \citet{Man19} and \citet{Man15}.
The microturbulent velocity for the target is fixed at the values of 0.5$\pm$0.5 km s$^{-1}$ for simplicity.

First, we use the FeH molecular lines in the Wing-Ford band at 990-1020 nm to estimate the effective temperature $T_{\rm eff}$. The band consists of more than 1000 FeH lines, and we select 47 lines with clear line profiles. The adopted data of the spectral line are available from the MARCS page.
We measure the EW of each FeH line with the Gaussian profile  and find the optimum $T_{\rm eff}$ so that the synthetic spectra reproduce the EW by an iterative search.
We note that we assume the solar values for the metallicity throughout this step. The average of estimated effective temperature $T_{\rm eff}$ for each of the 47 lines is set as the best value of $T_{\rm eff}$.
Its uncertainty is derived as the line-to-line scatter calculated by an estimated standard deviation from all the lines \citep{Ishikawa22}.

Second, adopting the value of $T_{\rm eff}$ estimated above, we derive the elemental abundances of Na, Mg, Ca, Ti, Cr, Mn, Fe, and Sr, from the corresponding atomic lines as in \citet{Ishikawa20}.
The data of the spectral line are also taken from the Vienna Atomic Line Database \citep{Kupka99,Ryabchikova15}.
We select the lines based on three criteria: not suffering from blending of other absorption lines; sensitive to elemental abundances; reasonably determined continuum level.
The EWs are measured by fitting synthetic spectra on a line-by-line basis.
We search for an elemental abundance until the synthetic EW corresponds to the observed one for each line and take the average for all the lines to estimate the abundance relative to hydrogen one as [X/H] for an element X.

Third, we adopt the iron abundance [Fe/H] above as the metallicity of the atmospheric model grid to recalculate $T_{\rm eff}$ as in the first step.
Then, we adopt the resultant $T_{\rm eff}$ to derive the elemental abundances with [Fe/H] as in the second step.
The procedure enables to derive $T_{\rm eff}$ and abundances until the convergence within the measurement errors.
As the final result of these analyses from the IRD spectra, we obtain $T_{\rm eff}$ = 3604 $\pm$ 101 K and $\lbrack{\rm Fe/H}\rbrack$ = 0.12 $\pm$ 0.18 dex for the host star TOI-654.
The effective temperature is derived to be consistent with the values in \citet{Hord24} ($3433$ K) and the TESS Input Catalog ($3424\pm157$ K) within two-sigma uncertainties.
The abundances for the other elements are listed in Table \ref{tb:stellar}.

\begin{figure}[tb]
\begin{center}
\includegraphics[width=8cm]{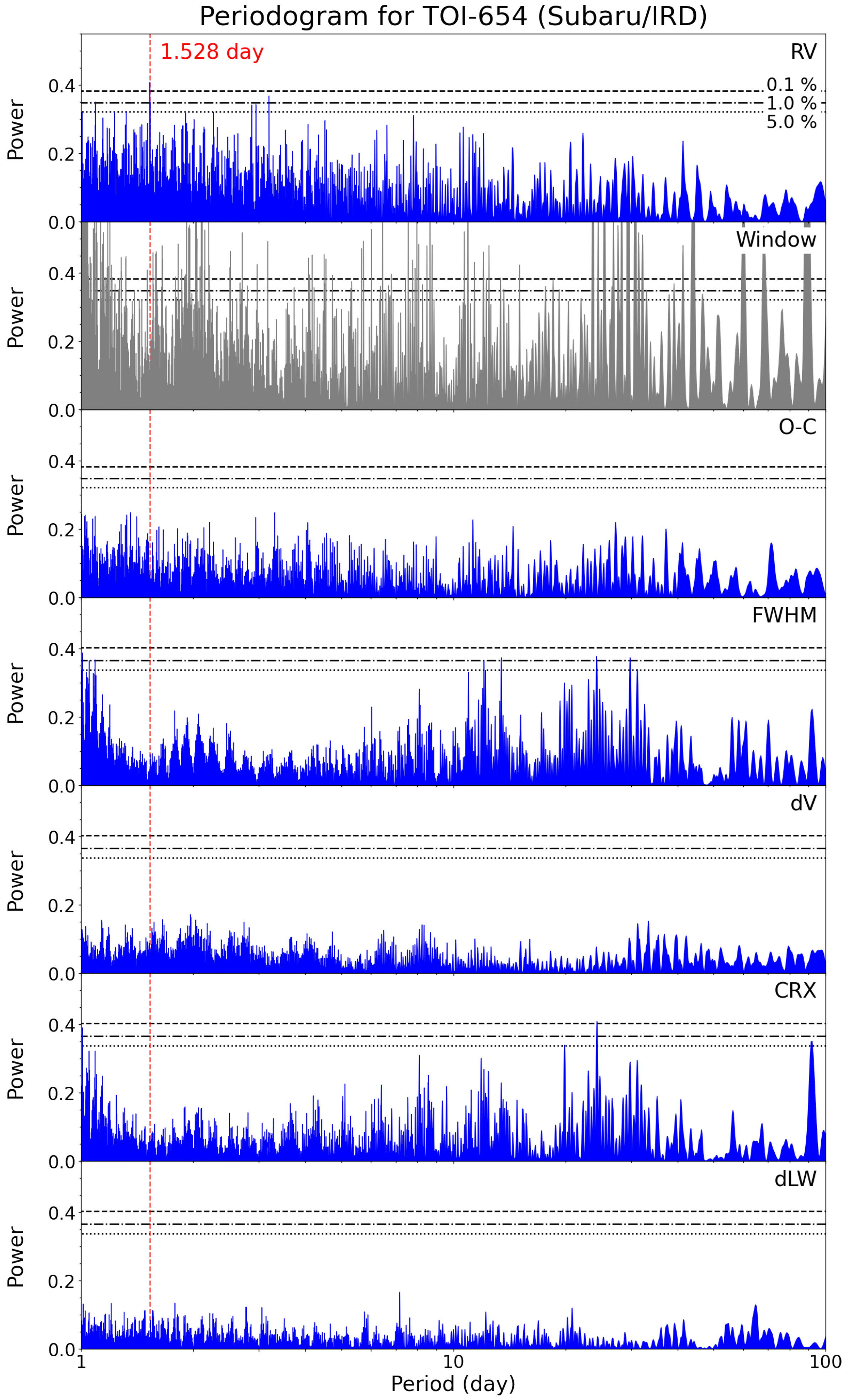}
\end{center}
\caption{Periodograms of the RV, Window, O-C, FWHM, dV, CRX, and dLW, obtained with the GLS from the IRD spectra for TOI-654 (Section \ref{sec:activity}).
The horizontal lines represent the FAP of 0.1, 1.0, and 5.0 \%, respectively (black) for each of the panel. In the top panel for the RV, the orbital period of the planet (=1.528 day) is detected with the FAP smaller than 0.1 \% (red).
{Alt text: Period (day) versus the power of the GLS periodogram for the RV, Window, O-C, FWHM, dV, CRX, and dLW, from the IRD spectra. In each panel, vertical line shows the orbital period of the transiting planet, and horizontal lines show the thresholds of the FAP of 0.1, 1.0, and 5.0 \%. In the top panel of the RV, the orbital period is detected with the FAP smaller than 0.1 \%.} 
} \label{fig:periodogram}
\end{figure}

\subsubsection{Photometric properties} \label{sec:sed}
We estimate and update the stellar parameters other than the metallicity from the photometric magnitudes.
First, we estimate the stellar radius and mass through their empirical relations with luminosity and metallicity \citep{Man15,Man19} by adopting parallax assumed to be inverse of the distance from Gaia DR3 \citep{Gaia}, $K_{\rm s}$-band magnitude from Two Micron All Sky Survey, and metallicity [Fe/H] from Section \ref{sec:metal}.
The stellar radii and mass are estimated to be $0.430 \pm 0.013$ $R_{\sun}$ and $0.419 \pm 0.009$ $M_{\sun}$, respectively.

Second, we derive the effective temperature $T_{\rm eff}$, stellar radius $R_{\rm s}$, and distance $d$, by fitting a stellar atmospheric model to the Spectrum energy distributions (SED) from the photometric magnitudes (Figure \ref{fig:SED}).
As the prior distributions for $T_{\rm eff}$ and $d$ to fit the SED, we adopt the normal distributions from the IRD spectra and Gaia DR3, respectively. We set an uniform prior distribution for $R_{\rm s}$. 
We derive posterior distribution of $T_{\rm eff}$, $R_{\rm s}$, and $d$, using a sampler of an affine invariant Markov chain Monte Carlo \texttt{emcee} \citep{emcee}.
As the photometric magnitudes, we use the magnitudes in the $BP$-, $G$-, $RP$-bands from Gaia DR3, $J$-, $H$-, $K_{\rm s}$-bands from 2MASS, and W1, W2 from AllWISE.
We linearly interpolate the tabulated BT-Settl model \citep{Allard14} to calculate a synthetic SED under certain values of $T_{\rm eff}$, and [Fe/H], and $\log g$. The values of [Fe/H] and $M_{\rm s}$ are sampled from normal distributions obtained from the IRD spectra and empirical relation. 
We add a parameter of the jitter error to all of the magnitude errors in the analysis because the errors are underestimated \citep{2MASS_error}.
As a result, the effective temperature $T_{\rm eff}$, radii $R_{\rm s}$, and distance $d$ are derived to be $3521 ^{+33}_{-48}$ K, $0.422 ^{+0.011}_{-0.009}$ $R_{\sun}$, and $57.851^{+0.055}_{-0.056}$ pc respectively. The stellar radius from the SED corresponds to that from the empirical relation within the uncertainty.
We note that the uncertainty of the effective temperature does not take into account the uncertainty inherent in the stellar atmospheric model for M dwarfs.
We adopt the value of $R_{\rm s}$ and $M_{\rm s}$ from the empirical relation in the first step, $T_{\rm eff}$ and $d$ from the SED in the second step, as the fiducial value of the stellar radius, mass, effective temperature, and distance, respectively (Table \ref{tb:stellar}). We also determine the spectrum type of TOI-654 to be M2V from $T_{\rm eff}$ and the difference of the magnitudes $J$-$H$ and $H$-$K$ based on \citet{Pecaut13}.

\subsubsection{Period analysis and stellar activity} \label{sec:activity}
We perform period analysis of the RV, RV after subtracting the planetary signal of the circular orbit (O-C) (Section \ref{sec:planet}), and activity indicators, FWHM, dV, CRX, and dLW (Section \ref{sec:ird}) with the Generalized Lomb-Scargle periodogram (GLS; \cite{GLS}).
In  Figure \ref{fig:periodogram}, we detect a periodic signal from the planet ($=1.528$ day) in the RV with a False Alarm Probability (FAP) less than 0.1\%. 
There are some periodic signals with the FAP < 1.0 \% in the FWHM and CRX, but these are in the window function.
Therefore, there are no significant periodic signals in the O-C and activity indicators.
We constrain the projected rotational velocity of TOI-654 to be $v \sin i_s < 3$ km s$^{-1}$ from the IRD spectra.

To investigate the stellar activity of the host star, we search for photometric variability in TESS light curve.
We calculate a GLS periodogram from the out-of-transit light curves, and we find no significant periodicity from the TESS light curve.
In addition, we search for periodic variability from the light curves of Zwicky Transient Facility (ZTF; \cite{ztf}) DR22 and the All-Sky Automated Survey for Supernovae (ASAS-SN; \cite{asas}).
We retrieve the ZTF light curves in $g$-, $r$-, and $i$-bands \citep{Masci19}, and the ASAS-SN light curves in V- and $g$-filters from the ASAS-SN portal \citep{Hart23}. We calculate a periodogram from each of the light curves after removing outliers. As results, we also find no significant periodicities of photometric variability due to stellar activity. The TESS, ZTF, and ASAS-SN light curves and their periodogram are shown in Appendix \ref{sec:activity_analysis}.

\begin{table*}[tb]
\caption{Parameters of TOI-654~b}
\begin{tabular}{lcc}
Parameter & Circular & Eccentric  \\ \hline
(Planetary parameters) & &   \\
$P$ (day) &$ 1.527561\pm0.000001$ & $ 1.527561\pm0.000001$
 \\
$T_{\rm c}$ (BJD-2457000) & $ 2243.0584 \pm 0.0003$ & $ 2243.0584\pm 0.0003$  \\
$b$  &$ 0.290 ^{+ 0.104}_{- 0.112}$& $ 0.391 ^{+ 0.186 }_{- 0.203}$   \\
$R_{\rm p}/R_{\rm s}$  &$ 0.0510 \pm 0.0008$ & $ 0.0517 \pm 0.0011$  \\
$\sqrt{e} \cos \omega$  &0 ({\rm fixed}) &   $ 0.130 ^{+ 0.129 }_{- 0.141 }$ \\
$\sqrt{e} \sin \omega$  & 0 ({\rm fixed}) & $ -0.140 ^{+ 0.229 }_{- 0.223 }$  \\
$K$ (m s$^{-1}$) &$ 8.62 ^{+ 1.22 }_{- 1.23 }$&  $ 8.86^{+ 1.25}_{- 1.23}$ \\ \hline
(RV parameters) & &     \\
$\gamma_{\rm IRD}$ (m s$^{-1}$) &$ 1.205 ^{+ 0.788}_{- 0.771}$ &  $ 1.079 ^{+ 0.834 }_{- 0.800 }$   \\ 
$\sigma_{\rm jit, IRD}$ (m s$^{-1}$)  & $ 4.06 ^{+ 1.08 }_{- 1.04 }$ ($<7.19$)$^\dag$& $ 3.74 ^{+ 1.30 }_{- 1.26 }$ ($<7.11$)$^\dag$   \\ \hline
$\log {\cal Z}$ & $-295.464$ & $-295.622$ \\ \hline \hline
(Derived parameters) & &     \\
$R_{\rm p}$ ($R_\earth$)&$  2.378\pm0.089$& $ 2.409 \pm  0.096$ \\
$M_{\rm p}$ ($M_\earth$)  &$ 8.71 \pm 1.25$ & $8.87 \pm 1.23$ \\
$a$ (au) &$0.01944 \pm 0.00015$ &  $0.01942 \pm 0.00015 $ \\
$e$ &$0$ ({\rm fixed})   & $ 0.097 ^{+ 0.077 }_{- 0.072}$ ($<0.373$)$^\dag$  \\
$\omega$ (deg) &$0$ ({\rm fixed}) & $ -25.63 ^{+ 64.89 }_{- 51.32 }$  \\
$i_{\rm p}$ (deg) &$88.28^{+ 0.71 }_{- 0.66}$  &$87.87 ^{+ 1.03 }_{- 0.88 }$    \\
$T_{14}$ (hour) & $1.200 \pm 0.013$ &  $ 1.209\pm0.019$ \\
$\rho_{\rm p}$ (g cm$^{-3}$)  &$3.59 \pm 0.65$ & $3.51\pm0.61$ \\ 
$\rho_{\rm p}$ ($\rho_{\earth,s}$)  &$0.42 \pm 0.08$ & $0.41\pm0.07$ \\ 
$T_{\rm eq}$ (K) ($A_B=0.0$)  &$794 \pm 15 $&  $795\pm15$  \\
$T_{\rm eq}$ (K) ($A_B=0.3$)  &$ 727 \pm 14$&  $727\pm14$  \\
$S$ ($S_{\earth}$) &$ 65.0 \pm 1.6$ &  $65.1 \pm1.6$ \\ 
TSM &$49\pm8$&   $50 \pm8$ \\
ESM &$11\pm1$& $11\pm1$  \\ \hline
\end{tabular}  \label{tb:para654}
\begin{tabnote}
\footnotemark[\dag] The upper limit of 99.73 \% confidence level.
\end{tabnote}
\end{table*}

\subsection{Planetary properties} \label{sec:planet}

\subsubsection{Transit model for photometry} \label{sec:tess_transit}
To deduce the parameters of the transiting planet, we fit the TESS, MuSCAT2, and MuSCAT3 light curves with the transit model \citep{Agol20} implemented in \texttt{jaxoplanet} \citep{jaxoplanet}.
The transit model is specified by the following parameters: the stellar mass $M_{\rm s}$, orbital period $P$, transit central time $T_{\rm c}$, 
impact parameter $b$, planetary radius relative to the stellar radius $R_{\rm p}/R_{\rm s}$, eccentricity $e$, argument of periastron $\omega$, and transformed quadratic limb-darkening coefficients $q_1$ and $q_2$ \citep{kipping13}.

The parameters $e$ and $\omega$ are transformed to $\sqrt{e} \sin \omega$ and $\sqrt{e} \cos \omega$ for their sampling efficiency \citep{Ford06}.
We calculate the values and uncertainties of $q_1$ and $q_2$ with \texttt{LDtk} \citep{ldtk} from the stellar parameters (Table \ref{tb:stellar}) and adopt normal prior distributions with three times larger uncertainties for these parameters because the systematic uncertainties could be inherent in the stellar model \texttt{PHOENIX} \citep{phoenix}. We adopt the transit ephemeris for $T_{\rm c}$ in the observation window of MuSCAT3. 
We set a normal prior distribution for $M_{\rm s}$ with the value from the empirical relation, a 
log-uniform prior distribution for $P$, and uniform prior distributions for other parameters. 

We include a jitter term $\sigma_{\rm jit}$ to account for the systematic noise in the TESS, MuSCAT2, and MuSCAT3 light curves for each of the band.
In the baseline model of the MuSCAT2 and MuSCAT3 light curves, we adopt the Gaussian process with the Mat\'{e}rn-3/2 kernel using \texttt{tinygp} \citep{tinygp} for each of the light curves.  The time-correlated systematic noise can be assumed to have a common time scale and be different only in the amplitude (e.g., \cite{Hayashi24}). The Mat\'{e}rn-3/2 kernel with the jitter kernel for the MuSCAT2 and MuSCAT3 light curves is described by 
\begin{align} \label{eq:kernel}
k(|t_{l,i}-t_{l,j}|) &= \sigma^2_{{\rm jit},l}\delta_{ij} \notag  \\
&+ \sigma_{{\rm sys},l}^2 \bigl( 1 + \frac{\sqrt{3} |t_{l,i}-t_{l,j}|}{\rho_{\rm sys}} \bigr) \notag\\ &\times \exp \bigl( -\frac{\sqrt{3} |t_{l,i}-t_{l,j}|}{\rho_{\rm sys}} \bigr),
\end{align}
where ($i$, $j$) and $l$ denote the numbers of data and label of band, respectively. We set log-uniform prior distributions for the jitter parameter $\sigma_{\rm jit}$ and GP hyperparameters $\sigma_{\rm sys}$ and $\rho_{\rm sys}$.
The data and optimum model are described in Figure \ref{fig:tess} (TESS) and \ref{fig:m2_654} (MuSCAT2 and MuSCAT3), from the joint fit with the RV data (Section \ref{sec:fit}).
The derived transit parameters are described in 
Table \ref{tb:para654}.
The stellar limb-darkening parameters and GP hyperparameters are listed in Table \ref{tb:add_para654}.

\subsubsection{Orbital model for radial velocity} \label{sec:rv}
\begin{figure}[tb]
\begin{center}
\includegraphics[width=8cm]{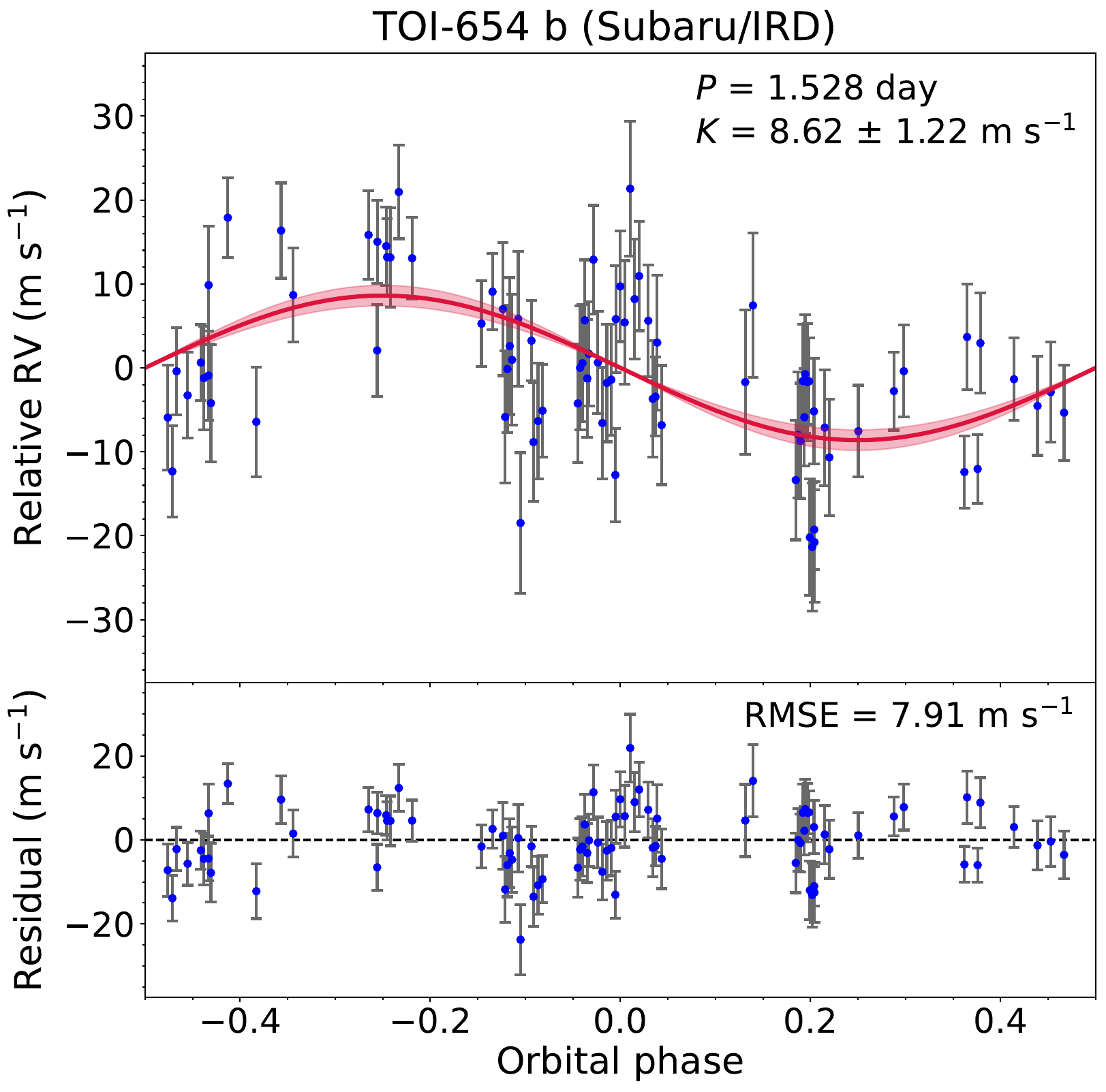}
\end{center}
\caption{The phase-folded RV data with the orbital period ($=1.528$ day) of the planet TOI-654~b (blue), the circular model with its uncertainty (red), and the residual from the model.
{Alt text: The orbital phase versus the relative RV (m s$^{-1}$) and residual (m s$^{-1}$). The relative RV is the RV minus offset, and the residual is the relative RV minus the optimum circular model.  The lines show the circular model and its uncertainty.} 
}\label{fig:fit_654}
\end{figure}

We fit the RV data from the IRD spectra with the orbital model in \texttt{jaxoplanet} \citep{jaxoplanet}. 
The RV model is specified by the following parameters: the orbital period $P$, transit central time $T_{\rm c}$, eccentricity $e$, argument of periastron $\omega$, and semi-amplitude $K$. In addition, we take account of the RV offset $\gamma_{\rm IRD}$ and the systematics with the jitter term $\sigma_{\rm jit, IRD}$.
The parameters of $P$, $T_{\rm c}$, $e$, and $\omega$ are common with the transit model for the planet TOI-654~b. We set a log-uniform prior distribution for $\sigma_{\rm jit, IRD}$ and uniform prior distributions for $K$ and $\gamma_{\rm IRD}$.
The data and optimum model are described in Figure \ref{fig:fit_654} from the joint fit with the transit data (Section \ref{sec:fit}), and
the derived RV parameters are described in 
Table \ref{tb:para654}.

\subsubsection{Deducing parameters and comparing models} \label{sec:fit}
To deduce the stellar and planetary parameters, we simultaneously fit the TESS, MuSCAT2 and MuSCAT3 light curves for the transiting planet TOI-654~b, and RV data using No U-turn Sampler \citep{NUTS} in \texttt{NumPyro} \citep{numpyro}.
In addition, to compare the models between circular and eccentric orbits in the Bayesian framework \citep{Nelson20}, we calculate the model evidence of the RV data for each of the models via importance sampling \citep{Diaz14, Ikuta20}. 
As a result, both of the model evidence are comparable given the difference of a logarithm of the model evidence $\Delta \log {\cal Z} \equiv \log {\cal Z}_{\rm circular}- \log {\cal Z}_{\rm eccentric}=  0.158$ \citep{Kass95}. Derived parameters from both models are almost consistent within their uncertainties, and the eccentricity in the eccentric model $e= 0.097^{+ 0.077 }_{- 0.072 }$ is almost consistent with that of the circular model. The parameter uncertainties in the eccentric model are slightly larger due to taking account of the eccentricity $e$ and argument of periastron $\omega$. In Section \ref{sec:discussion}, we discuss the planetary compositions and atmospheric observations mainly based on the circular model given its slightly larger model evidence.

\begin{figure}[tb]
\begin{center}
\includegraphics[width=8cm]{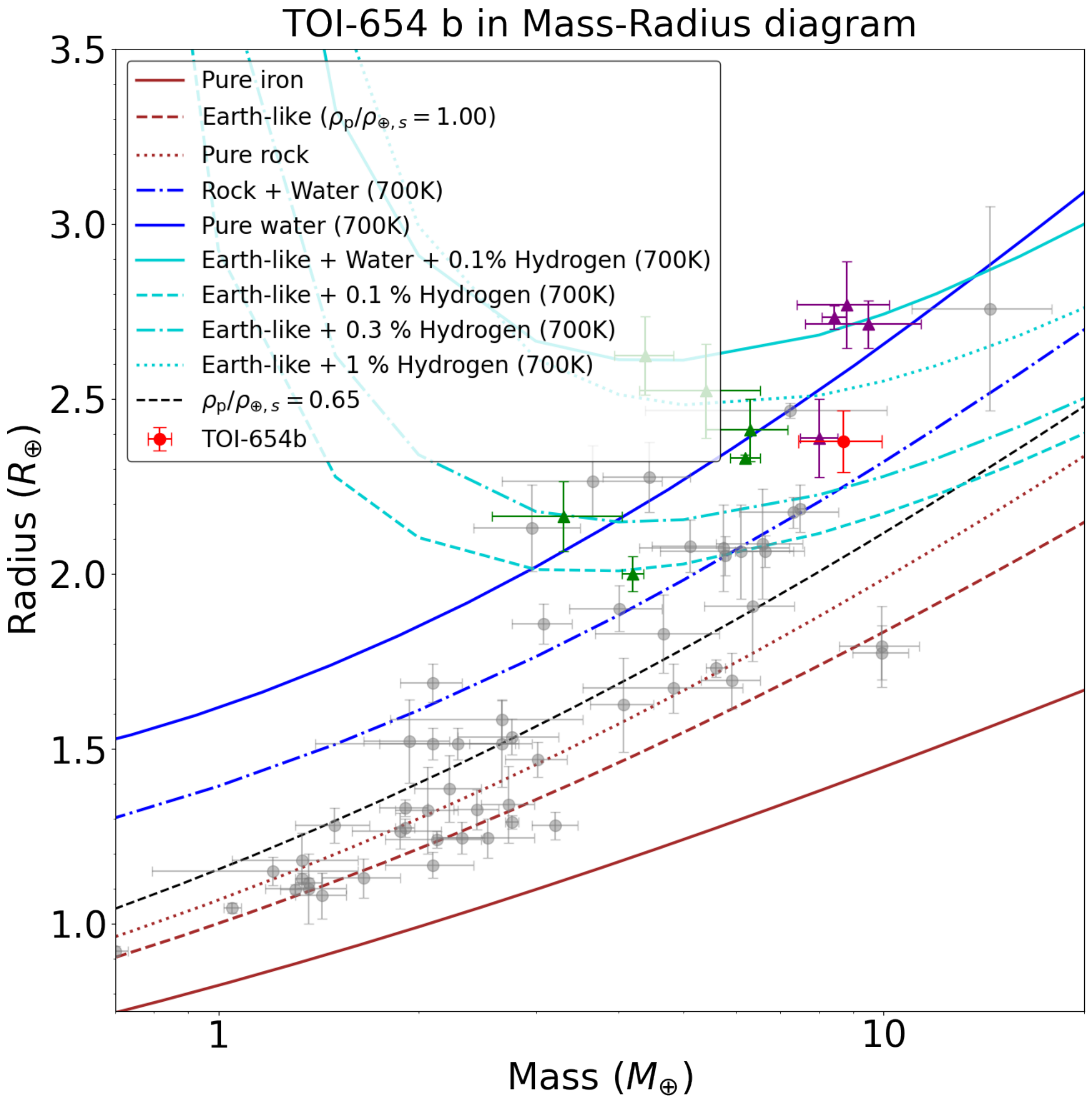}
\end{center}
\caption{Mass-Radius diagram for TOI-654~b (red) in comparison with other well-characterized small planets around M dwarfs ($M_{\rm p}\leq 20 M_{\earth}$, $R_{\rm p}\leq 3 R_{\earth}$, $T_{\rm eff}\leq 4000$ K, and the mass and radius divided by their uncertainties are less than 40\%) and various composition models \citep{Zeng19}: Pure iron, Earth-like, pure rock (brown), rock and water, pure water (blue), rock and water with 0.1\% hydrogen, and rock with 0.1\%, 0.3\% hydrogen, and 1\% hydrogen, under the temperature of 700 K at the surface pressure of 1 mbar (cyan), respectively. The black line is shown for the threshold of the water world ($\rho_{\rm p}/\rho_{\earth,s}=0.65$) proposed in \citet{Luque22}.
Small planets with mass comparable to TOI-654 and most suitable for the atmospheric observations are marked in purple (TOI-732c, GJ1214b, K2-18b, and TOI-269b) and other sub-Neptunes with the TSM > 90 are marked in green. 
{Alt text: The diagram of the mass ($M_\earth$) versus radius ($R_\earth$) for small planets around M dwarfs. The lines show various composition models. The diamond shows the transiting planet TOI-654~b in this paper. The triangle shows sub-Neptunes suitable for the atmospheric observations.
} 
} \label{fig:MR_654}
\end{figure}

\begin{figure}[tb]
\begin{center}
\includegraphics[width=8cm]{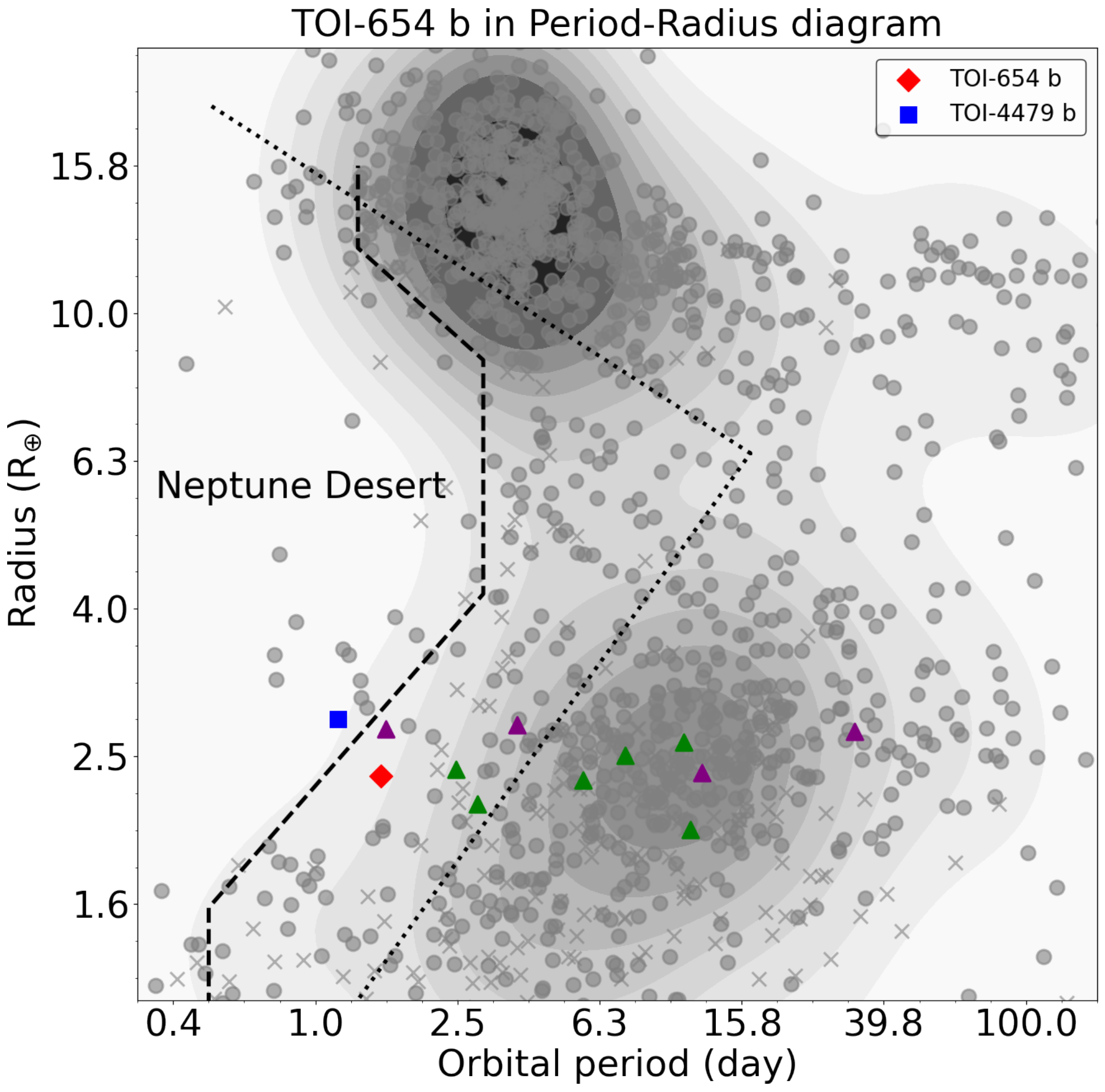}
\end{center}
\caption{Period-Radius diagram for TOI-654~b (red), TOI-4479~b (blue), and small planets suitable for the atmospheric observations (green and purple in Figure \ref{fig:MR_654}), in comparison with other planets (gray circle and cross for $T_{\rm eff}> 4000$ K and $T_{\rm eff} \leq 4000$ K, with the radius divided by their uncertainties are less than 40\%, respectively). 
TOI-654~b is located in the original Neptune desert (dotted black; \cite{Mazeh16}) and on the boundary of the Neptune desert from the recent observation (dashed black; \cite{2024CastroGonzalez}).
{Alt text: The diagram of the period (day) versus radius ($R_\earth$).
The diamond shows the transiting planet TOI-654~b in this paper. TOI-654~b is located in or on the boundary of the Neptune desert. The square shows a shorter-period sub-Neptune around M-dwarf.
The triangle shows sub-Neptunes suitable for the atmospheric observations.
} 
} \label{fig:desert}
\end{figure}

\section{Discussion} \label{sec:discussion}
\subsection{Planetary composition}
The mass is derived to be $M_{\rm p} = 8.71 \pm 1.25  M_{\earth}$, and the radius is updated to be $R_{\rm p} = 2.378 \pm 0.089 R_{\earth}$, from the joint analysis of the transit and RV data with the circular model. 
The precision of the planetary radius is improved from 5.3 \% ($R_{\rm p} = 2.371 \pm 0.125  R_{\earth}$ in \cite{Hord24}) to 3.7 \% because of the precise measurement of the stellar radius and additional TESS, MuSCAT2, and MuSCAT3 data.
We derive the mean density $\rho_{\rm p}=3.59 \pm 0.65 $ g cm$^{-3}$, and the equilibrium temperature assuming albedo values of 0.0 and 0.3 are also calculated to be $T_{\rm eq} = 794\pm15$ and $727\pm14$ K, respectively.
In Figure \ref{fig:MR_654}, we compare the derived mass and radius of TOI-654~b with various composition models \citep{Zeng19} and other small planets around M dwarfs ($M_{\rm p}\leq 20 M_{\earth}$, $R_{\rm p}\leq 3 R_{\earth}$, and $T_{\rm eff}\leq 4000$ K), whose planetary mass and radius are both measured with precisions better than 40 \% from the TEPcat database \citep{Southworth11}.
TOI-654~b is equally close to the composition models of 
an Earth-like core (32.5 wt\% Fe and 67.5 wt\% MgSiO3) with 0.3 wt\% hydrogen envelope and
rock with rich water (50 wt\% Earth-like rocky core and 50 wt\% H$_2$O layer) under the temperature of 700 K at the surface pressure of 1 mbar.
The mean density is normalized to be $\rho_{\rm p}/\rho_{\earth,s} = 0.42 \pm 0.08$ by the Earth-like composition in \citet{Zeng19}.
The deduced density can be mainly explained by a rocky core with a small amount of H/He envelope (e.g., \cite{Rogers23}).
On the other hand, TOI-654~b is located in the region of the water world proposed in \citet{Luque22}.
It is theoretically predicted that short-period sub-Neptunes around M dwarfs would accrete much water \citep{Kimura22}. Thus, future atmospheric observations would be required to verify the degenerated compositions (Section \ref{sec:atmos}).

The mass and radius of TOI-654~b are comparable with those of 
TOI-732~c (LTT 3780~c; \cite{Cloutier20b, Nowak20, Bonfanti24}), TOI-1470~b, TOI-1470~c \citep{Alvarez23}, and K2-146~c \citep{Hamann19, Lam20} among well-characterized small planets around M dwarfs. 
In addition, the mass of TOI-654~b is comparable to those of GJ1214~b (e.g., \cite{Charbonneau09, Cloutier21,Mahajan24}), K2-18~b \citep{Cloutier17, Sarkis18, Benneke19}, and TOI-269~b \citep{Cointepas21}, but the radius is smaller than those.
This difference in the radius would be explained by differences in the ratio of internal compositions.
The mass measurement of more super-Earths and sub-Neptunes around M dwarfs is required to verify the radius valley and water world scenario through survey programs \citep{Lacedelli24}.

In Figure \ref{fig:desert}, we compare the period and radius of TOI-654~b with other planets whose radius is measured with precisions better than 40 \% in the TEPcat database \citep{Southworth11}. TOI-654~b is also located on the outer edge in orbital period and lower edge in radius of the original Neptune desert \citep{2011SzaboKiss,Mazeh16, Lopez17} and the boundary of the recent observed one \citep{2024CastroGonzalez}, whose scarcity of planets is explained by the photoevaporation \citep{Owen17}. 
Recently, some sub- or super-Neptunes around M dwarfs have been discovered on and near the desert (e.g., \cite{Murgas21,Mori22, 2024PelaezTorres}).
In particular, only one sub-Neptune TOI-4479~b ($R_{\rm p} = 2.82 ^{+ 0.65}_{- 0.63} R_{\earth}$, and $P=1.16$ days) has shorter period than that of TOI-654~b among sub-Neptunes around M dwarfs as of February, 2025.
Thus, TOI-654~b is one of unique planets that allows us to constrain the formation process from both points of the radius valley and Neptune desert around M dwarfs.

\subsection{Future atmospheric observations} \label{sec:atmos}
TOI-654~b is validated as a suitable target for the atmospheric observation with the JWST \citep{Hord24} and included in the target list of the Ariel \citep{Edwards22}. 
We calculate the transmission and emission spectroscopy metric (TSM and ESM; \cite{Kempton18}) of TOI-654~b 
from the planetary mass and radius for the transmission and emission spectroscopy with the JWST and Ariel.
The TSM and ESM are updated to be $49\pm8$ and $11\pm1$ from the circular model (Table \ref{tb:para654}) while the values are previously calculated to be 78 and 9 \citep{Hord24} using the empirical radius-mass relation \citep{Chen17}.
Although the TSM of TOI-654~b is smaller than the threshold of the TSM $>90$  in the radius range of $1.5 < R_{\rm p}/ R_{\earth} < 10$
for the transmission spectroscopy \citep{Kempton18}, \citet{Hord24} suggests that TOI-654~b is one of suitable targets for the observation especially with the emission spectroscopy.
In the eccentric model, we derive the time difference between the transit and secondary eclipse from the circular model \citep{Win10}
and the eclipse duration for the emission spectroscopy to be $\Delta T_{\rm eccentric} - 
\Delta T_{\rm circular} \simeq P e \cos \omega (4-3 e \sin \omega)/ (2 \pi)  =  1.122^{+ 1.291 }_{- 1.149} $ and $T_{14,{\rm eclipse}}=  1.089 ^{+ 0.149 }_{- 0.189}$ hours.
Both from the transmission and emission spectroscopy in the precise ephemeris, we would expect to distinguish degenerated compositions of a rocky and volatile-rich core or rocky core surrounded by a H/He envelope (e.g., \cite{Cadieux24,Benneke24}).
We can also investigate atmospheric properties of the haze and metallicity (e.g., \cite{Kempton23,Gao23, Brande24}). 

\section{Conclusion} \label{sec:conclusion}
We report on follow-up observations of a planetary system with a transiting short-period sub-Neptune around the mid-M dwarf TOI-654. The transiting planet TOI-654~b is validated as a sub-Neptune with an orbital period of $P=1.53$ day as a suitable target for the atmospheric observation.
We measure the planetary mass and stellar properties with the IRD mounted on the Subaru telescope and obtain the stellar and planetary properties from additional transit observations by the TESS and ground-based multicolor photometry with MuSCAT2 and MuSCAT3. 
The planetary mass of TOI-654~b is determined to be $M_{{\rm p}} = 8.71 \pm 1.25 M_{\earth}$ from the RV data, and the radius is updated to be $R_{\rm p} =  2.378 \pm  0.089 R_{\earth}$ from the transit data. The bulk density suggests that the planet is composed of a rocky and volatile-rich core or a rocky core surrounded by a small amount of H/He envelope. 
TOI-654~b is one of unique planets located not only around the radius valley and but on the outer edge of the Neptune desert, and the precise mass determination enables us to constrain the atmospheric properties with future spectroscopic observations especially for the emission by the JWST and Ariel.

\begin{ack}
We sincerely appreciate the anonymous referee for providing careful feedback that helped to improve both content of this manuscript
and the clarity. This research is mainly based on data collected at the Subaru Telescope, which is operated by the National Astronomical Observatory of Japan. We are honored and grateful for the opportunity of observing the Universe from Maunakea, which has the cultural, historical and natural significance in Hawaii.
This paper is based on observations with the MuSCAT2 instrument, developed by Astrobiology Center, at Telescopio Carlos S\'{a}nchez operated on the island of Tenerife by the IAC in the Spanish Observatorio del Teide.
This paper is also based on observations with the MuSCAT3 instrument, developed by Astrobiology Center and under financial supports by JSPS KAKENHI (JP18H05439) and JST PRESTO (JPMJPR1775), at Faulkes Telescope North on Maui, HI, operated by the Las Cumbres Observatory.
This study was partly supported by the JSPS KAKENHI Grant Numbers JP15H02063, JP16K17660, JP17H01153, JP17H02871, JP17H04574, JP18H01265, JP18H05439, JP18H05442, JP19KK0082, JP19K14783, JP20H00170, JP20K14518, JP20K14521, JP21H00035, JP21H01141, JP21K13955, JP21K13975, JP21K13987, JP21K18638, JP21K20376, JP21K20388, JP22J11725, JP23H00133, JP23H01224, JP23H01227, JP23K17709, JP23K25920, JP23K25923, JP24H00017, JP24H00242, JP24H00248, JP24K00689, JP24K07108, JP24K17082, JP24K17083, JSPS Bilateral Program Number JPJSBP120249910, JSPS Grant-in-Aid for JSPS Fellows Grant Number JP20J21872, JP22KJ0816, JP24KJ0241, Astrobiology Center SATELLITE Research project AB022006, JST SPRING Grant Number JPMJSP2108, and JST CREST Grant Number JPMJCR1761.
F.M. acknowledges the financial support from the Agencia Estatal de Investigaci\'{o}n del Ministerio de Ciencia, Innovaci\'{o}n y Universidades (MCIU/AEI) through grant PID2023-152906NA-I00.
\end{ack}

\appendix 

\section{TESS, ZTF, and ASAS-SN light curves and their periodogram} \label{sec:activity_analysis}
\begin{figure*}[htb]
\begin{center}
\includegraphics[width=16cm]{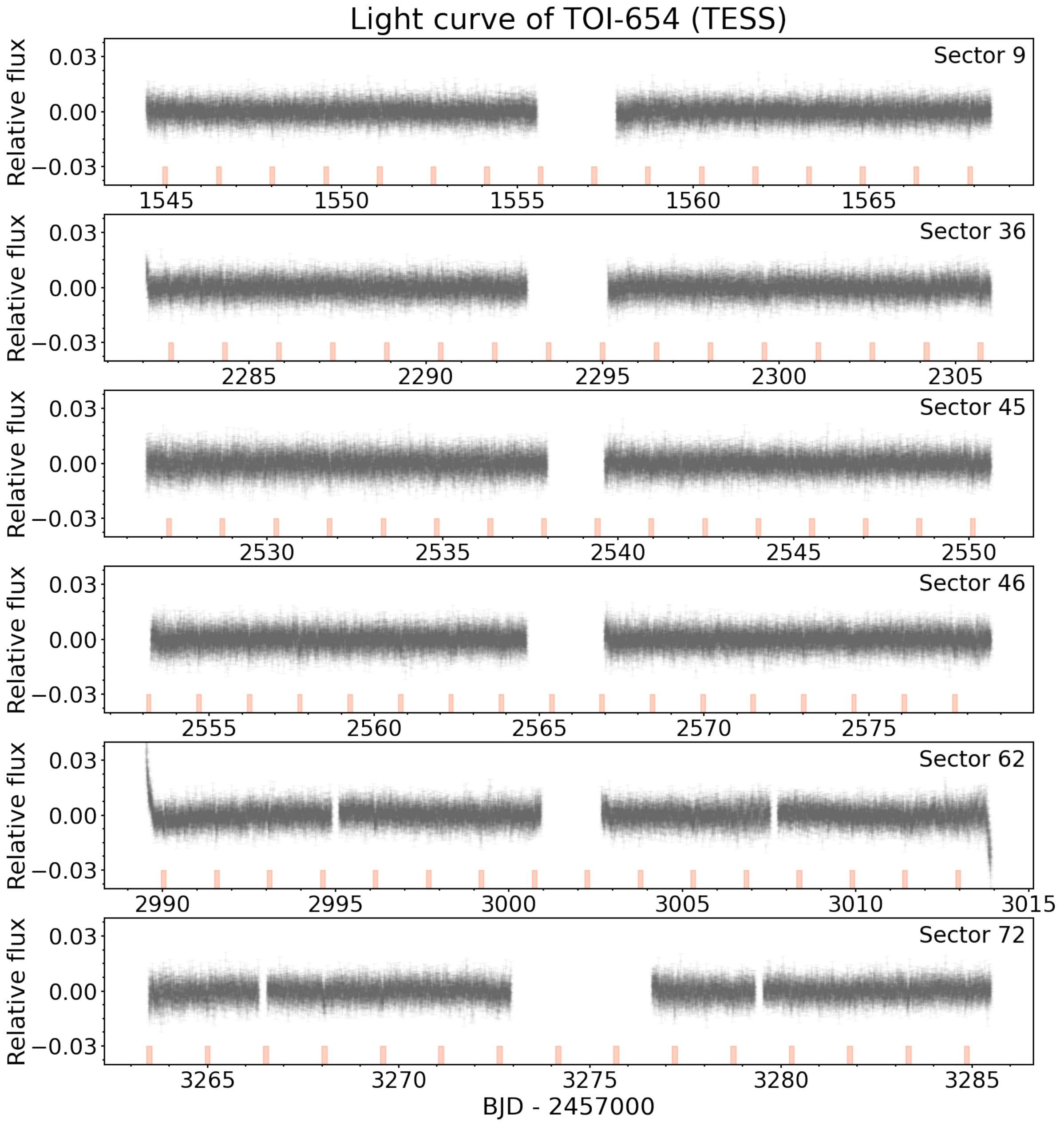}
\end{center}
\caption{
TESS light curves of TOI-654 from the PDC-SAP flux in its sectors 9, 36, 45, 46, 62, and 72 (Section \ref{sec:tess}). The transit windows are marked in orange (Figure \ref{fig:tess}). 
{Alt text: Time-series flux for all the TESS data with the transit windows.
}
}
\label{fig:tess_lc}
\end{figure*}

\begin{figure*}[htb]
\begin{center}
\includegraphics[width=16cm]{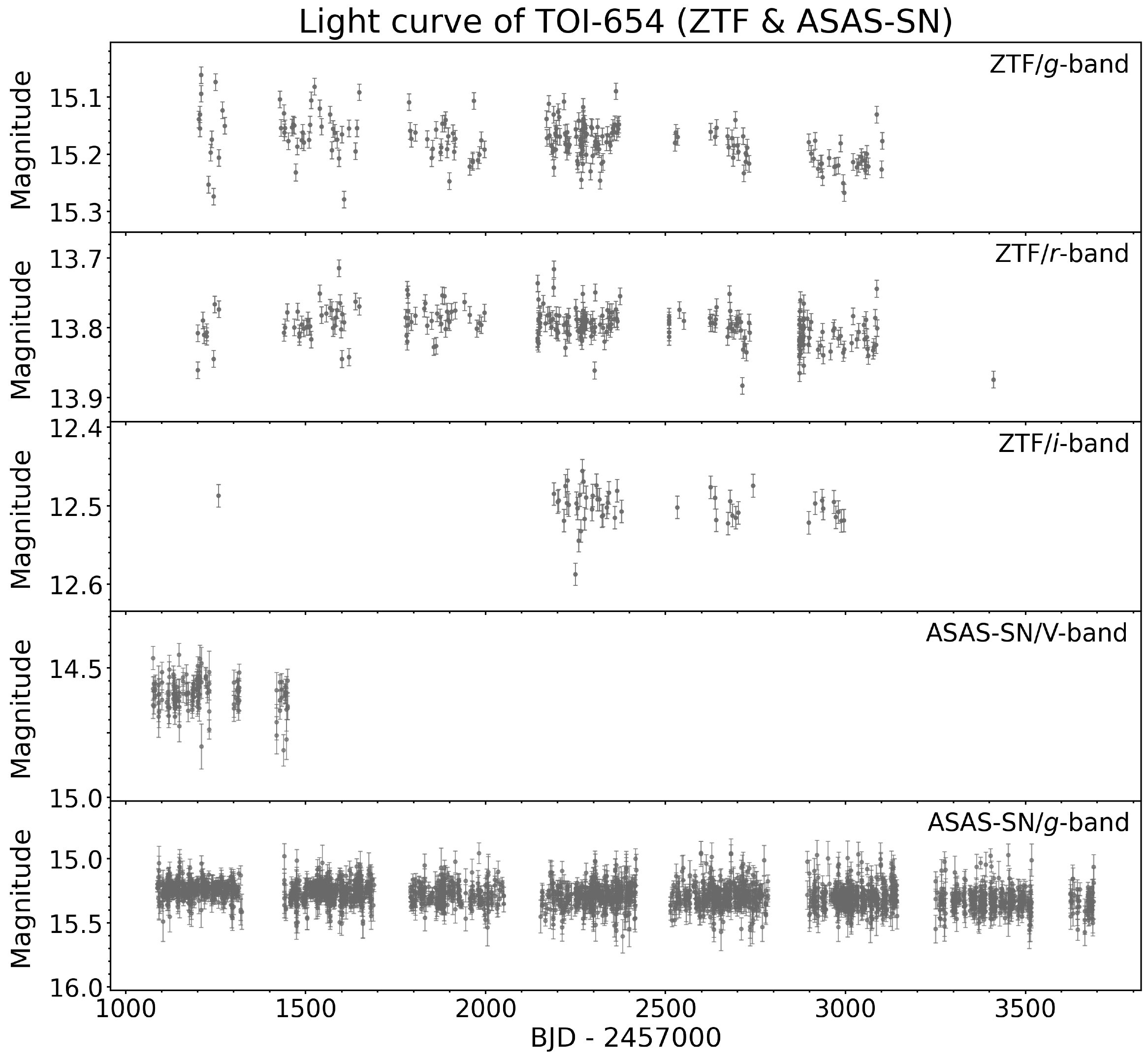}
\end{center}
\caption{
The ZTF light curves in $g$-, $r$-, and $i$-bands, and ASAS-SN light curves in V- and $g$-bands (Section \ref{sec:activity}).
{Alt text: Time-series flux for the ZTF and ASAS-SN data.
} 
}
\label{fig:ztf_lc}
\end{figure*}

\begin{figure*}[htb]
\begin{center}
\includegraphics[width=12cm]{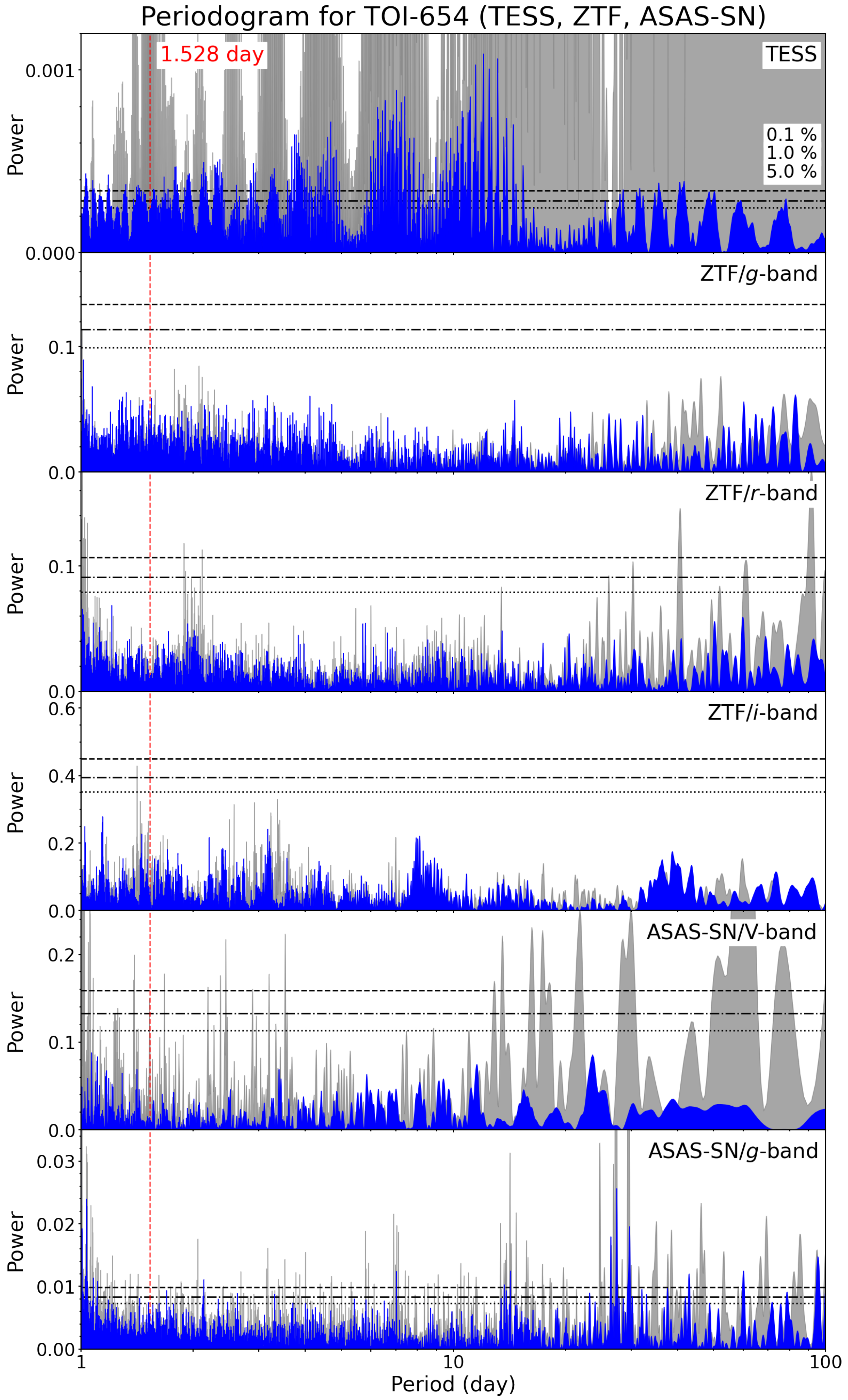}
\end{center}
\caption{
Periodograms for the TESS light curves, ZTF light curves in $g$-, $r$-, and $i$-bands, and ASAS-SN light curves in V- and $g$-bands (blue), and their window functions (gray), with the GLS for TOI-654 (Section \ref{sec:activity}).
In each of the panel, the vertical line represent the orbital period (=1.528 day) of the planet (red), and the horizontal lines represent the FAP of 0.1, 1.0, and 5.0 \%, respectively (black).
{Alt text: Period (day) versus the power of the GLS periodogram for the TESS, ZTF, and ASAS-SN light curves. In each panel, vertical line shows the orbital period of the planet, and horizontal lines show the thresholds of the FAP of 0.1, 1.0, and 5.0 \%. }  
}
\label{fig:ztf_period}
\end{figure*}

Figure \ref{fig:tess_lc} shows the TESS light curves in its sectors 9, 36, 45, 46, 62, and 72 (Section \ref{sec:tess}), and Figure \ref{fig:ztf_lc} shows the ZTF light curves in $g$-, $r$-, and $i$-bands, and ASAS-SN light curves in V- and $g$-bands. There is no significant photometric variability ascribed to stellar activity. Figure \ref{fig:ztf_period} shows the GLS periodogram for each of the light curve (Section \ref{sec:activity}).

\section{Other parameters for the transit data} \label{sec:appendix}
Table \ref{tb:add_para654} lists derived limb-darkening parameters for the TESS, $g$-, $r$-, $i$-, and $z_s$-bands, and hyperparameters for the baseline model of MuSCAT2 and MuSCAT3 data with the Gaussian process (Section \ref{sec:tess_transit}).

\bibliography{ird}

\begin{table*}[htb]
\caption{Other parameters of TOI-654~b}
\begin{tabular}{lcc}
Parameter &
Circular &Eccentric  \\ \hline
(\textit{Limb-darkening coefficients}) & &    \\
$q_{1,{\rm TESS}}$ & $ 0.361\pm 0.006$ &  $ 0.362\pm 0.006$  \\ 
$q_{2,{\rm TESS}}$ & $ 0.232\pm 0.006$ &  $ 0.231\pm 0.006$ \\ 
$q_{1,g}$ & $0.683\pm 0.010$ &    $0.684\pm 0.010$ \\ 
$q_{2,g}$ & $0.323\pm 0.008$ & $0.324\pm 0.008$  \\ 
$q_{1,r}$ & $0.576\pm 0.007$ & $0.575\pm 0.007$   \\ 
$q_{2,r}$ & $0.354\pm 0.008$ &  $0.354\pm 0.008$ \\ 
$q_{1,i}$ & $0.421 \pm 0.008$ &  $0.420 \pm 0.008$  \\ 
$q_{2,i}$ & $0.241 \pm 0.008$& $0.241\pm 0.007$   \\ 
$q_{1,z_s}$ &$0.324 \pm 0.008$ &  $0.324\pm 0.008$  \\ 
$q_{2,z_s}$ &$0.209 \pm 0.008$ & $0.209 \pm 0.008$    \\ 
\hline 
(\textit{GP parameters})$^\dag$ & &   \\ 
$\log \sigma_{{\rm jit},{\rm TESS}}$ & $ -11.906 ^{+ 2.179 }_{- 2.195}$ & $ -11.893 ^{+ 2.170 }_{- 2.145}$  \\ 
$\log \sigma_{{\rm jit},{\rm M2}, g}$ &$ -11.257^{+ 2.672 }_{- 2.621}$ &  $ -11.170^{+ 2.602 }_{- 2.498 }$ \\ 
$\log \sigma_{{\rm jit},{\rm M2},r}$ &  $ -10.009 ^{+ 2.598 }_{- 3.202 }$
& $ -9.681 ^{+ 2.305}_{- 2.977}$   \\ 
$\log \sigma_{{\rm jit},{\rm M2},i}$ &$ -8.170 ^{+ 1.022 }_{- 1.079 }$&  $ -8.043 ^{+ 0.900 }_{- 0.625 }$  \\ 
$\log \sigma_{{\rm jit},{\rm M2},z_s}$ & $ -6.972^{+ 0.134 }_{- 0.130 }$ & $ -6.974 ^{+ 0.136 }_{- 0.135 }$  \\ 
$\log \sigma_{{\rm sys},{\rm M2},g}$ &$ -6.643 ^{+ 0.360 }_{- 0.366}$& $ -6.678 ^{+ 0.369 }_{- 0.361 }$  \\ 
$\log \sigma_{{\rm sys},{\rm M2},r}$ & $ -6.446 ^{+ 0.271 }_{- 0.273 }$&  $ -6.442 ^{+ 0.282 }_{- 0.278 }$ \\ 
$\log \sigma_{{\rm sys},{\rm M2},i}$ &$ -6.196^{+ 0.262 }_{- 0.266 }$ & $ -6.199^{+ 0.275 }_{- 0.269 }$  \\ 
$\log \sigma_{{\rm sys},{\rm M2},z_s}$ &  $ -6.420^{+ 0.279 }_{- 0.282 }$ & $ -6.434 ^{+ 0.275 }_{- 0.263 }$ \\ 
$\log \rho_{\rm sys, M2}$ & $ -3.634^{+ 0.231 }_{- 0.229 }$ &  $ -3.635 ^{+ 0.214 }_{- 0.218 }$ \\  

$\log \sigma_{{\rm jit},{\rm M3}, g}$ & $ -10.680^{+ 2.891}_{- 2.911 }$ & $ -10.814^{+ 2.974 }_{- 3.022}$ \\ 
$\log \sigma_{{\rm jit},{\rm M3},r}$ & $ -8.171^{+ 1.470}_{- 2.747}$ & $ -8.672 ^{+ 1.956}_{- 3.826}$ \\ 
$\log \sigma_{{\rm jit},{\rm M3},i}$ & $ -8.306 ^{+ 1.179 }_{- 1.868}$ & $ -8.400 ^{+ 1.276 }_{- 2.163}$   \\ 
$\log \sigma_{{\rm jit},{\rm M3},z_s}$ &$ -11.538^{+ 2.277 }_{- 2.287 }$&  $ -11.876 ^{+ 2.395 }_{- 2.303 }$ \\ 
$\log \sigma_{{\rm sys},{\rm M3},g}$ &$ -5.805^{+ 0.512}_{- 0.508}$ & $ -5.668 ^{+ 0.581}_{- 0.575 }$   \\ 
$\log \sigma_{{\rm sys},{\rm M3},r}$ &$ -6.808^{+ 0.438}_{- 0.435 }$& $ -6.694 ^{+ 0.510 }_{- 0.493 }$   \\ 
$\log \sigma_{{\rm sys},{\rm M3},i}$ &$ -7.367^{+ 0.399 }_{- 0.402 }$& $ -7.401 ^{+ 0.406 }_{- 0.399 }$    \\ 
$\log \sigma_{{\rm sys},{\rm M3},z_s}$ &$ -7.005^{+ 0.484 }_{- 0.496 }$ &$ -7.002^{+ 0.483}_{- 0.481}$   \\ 
$\log \rho_{\rm sys, M3}$ & $ -3.063^{+ 0.653}_{- 0.652 }$&  $ -2.908 ^{+ 0.678 }_{- 0.720}$  \\ 
\hline
\end{tabular}\label{tb:add_para654} 
\begin{tabnote}
\par\noindent
\footnotemark[\dag] The GP kernel is given by Equation \ref{eq:kernel} for MuSCAT2 (M2) and MuSCAT3 (M3) data.
\par\noindent
\end{tabnote}
\end{table*}

\bibliographystyle{aasjournal}
\end{document}